\documentclass[sigconf]{acmart}

\usepackage{cleveref}

\usepackage{algorithm}
\usepackage{algpseudocode} 

\usepackage{makecell}

\AtBeginDocument{%
  }

\usepackage{CJKutf8}

\copyrightyear{2026}
\acmYear{2026}
\setcopyright{cc}
\setcctype{by}
\acmConference[CHI '26]{Proceedings of the 2026 CHI Conference on Human Factors in Computing Systems}{April 13--17, 2026}{Barcelona, Spain}
\acmBooktitle{Proceedings of the 2026 CHI Conference on Human Factors in Computing Systems (CHI '26), April 13--17, 2026, Barcelona, Spain}
\acmPrice{}
\acmDOI{10.1145/3772318.3791152}
\acmISBN{979-8-4007-2278-3/2026/04}

\begin{document}

\begin{CJK*}{UTF8}{gbsn}

\title{Tower of Babel in Cross-Cultural Communication: A Case Study of \#Give Me a Chinese Name\# Dialogues During the ``TikTok Refugees'' Event}

\author{Jielin Feng}
\email{jielinfeng23@m.fudan.edu.cn}
\orcid{0009-0008-1943-5609}
\affiliation{%
  \institution{School of Data Science, Fudan University}
  \city{Shanghai}
  \country{China}
}

\author{Zhibo Yang}
\email{22300130030@m.fudan.edu.cn}
\orcid{0009-0006-9679-5101}
\affiliation{%
  \institution{School of Journalism, Fudan University}
  \city{Shanghai}
  \country{China}
}

\author{Jingyi Zhao}
\email{22300120121@m.fudan.edu.cn}
\orcid{0009-0006-7355-5519}
\affiliation{%
  \institution{School of Foreign Language and Literature, Fudan University}
  \city{Shanghai}
  \country{China}
}

\author{Yujia Li}
\email{23301170062@m.fudan.edu.cn}
\orcid{0009-0003-8846-8889}
\affiliation{%
  \institution{School of Journalism, Fudan University}
  \city{Shanghai}
  \country{China}
}

\author{Xinwu Ye}
\email{xinwuye43@connect.hku.hk}
\orcid{0009-0008-3164-0373}
\affiliation{%
  \institution{School of Data Science, Fudan University}
  \city{Shanghai}
  \country{China}
}
\affiliation{%
  \institution{School of Computing and Data Science, The University of Hong Kong}
  \city{Hong Kong}
  \country{China}
}

\author{Xingyu Lan}
\authornote{Xingyu Lan and Siming Chen are co-corresponding authors. Xingyu Lan is also a member of the Research Group of Computational and AI Communication at the Institute for Global Communications and Integrated Media.}
\email{xingyulan96@gmail.com}
\orcid{0000-0001-7331-2433}
\affiliation{%
  \institution{School of Journalism, Fudan University}
  \city{Shanghai}
  \country{China}
}

\author{Siming Chen}
\authornotemark[1]
\email{simingchen@fudan.edu.cn}
\orcid{0000-0002-2690-3588}
\affiliation{%
  \institution{School of Data Science, Fudan University}
  \city{Shanghai}
  \country{China}
}
\affiliation{%
  \institution{Qingdao Research Institute, Fudan University}
  \city{Shanghai}
  \country{China}
}

\renewcommand{\shortauthors}{Feng et al.}

\renewcommand{\shorttitle}{Tower of Babel in Cross-Cultural Communication: A Case Study of \#Give Me a Chinese \\Name\# Dialogues During the ``TikTok Refugees'' Event}

\renewcommand{\sectionautorefname}{Section}
\renewcommand{\subsectionautorefname}{Section}
\renewcommand{\subsubsectionautorefname}{Section}

\begin{abstract}
The sudden influx of ``TikTok refugees'' into the Chinese platform RedNote in early 2025 created an unprecedented, large-scale online cross-cultural communication event between the West and East. Although prior HCI research has studied user behavior in social media, most work remains confined to monolingual or single-cultural contexts, leaving cross-linguistic and cultural dynamics underexplored. To address this gap, we focused on a particularly challenging cross-cultural encoding–decoding task that remains stubbornly beyond the reach of machine translation, i.e., foreign newcomers asking Chinese users for Chinese names, and examined how people collectively constructed a digital ``Babel Tower'' through various information encoding strategies. We collected and analyzed over 70,000 comments from RedNote with a creative human-in-the-loop approach using large language models, deriving a systematic framework summarizing cross-cultural information encoding strategies, how they are combined and layered to complicate decoding, and how they relate to \textcolor{black}{engagement metrics such as the number of likes}. 

\end{abstract}



\maketitle
\section{Introduction}
In early 2025, ``TikTok refugees'' became one of the most trending and influential social-media topics. Triggered by renewed calls to ban TikTok in the United States over national security concerns, more than three million self-described English-speaking users flocked to a Chinese social media platform \cite{liu2025tiktok}, RedNote\footnote{\url{https://www.xiaohongshu.com/}}, \textcolor{black}{and discussions around this incident accumulated over 2.45 billion views \cite{yuan2025love}.}
As one of the most popular Chinese social media platforms, RedNote is well-known for its combination of lifestyle sharing and community interaction~\cite{huang2024domesticating}. \textcolor{black}{Similar to Instagram, it centers on text- and image-based posts with rich captions and threaded comments, which makes it a particularly suitable venue for dialogic exchanges between users.}
\textcolor{black}{Despite the plethora of existing social media platforms in China (e.g., Weibo, Douyin), RedNote surprisingly emerged as the central hub and symbolic platform of the exodus, becoming the primary destination for direct encounters between Western newcomers and Chinese users.}

In the field of human-computer interaction (HCI), \textcolor{black}{prior research concerning social media has examined various platforms and user behaviors, such as danmaku and forum comments on Bilibili~\cite{wu2018danmaku}, hashtag use on RedNote~\cite{wan2025hashtag}, YouTube addiction~\cite{de2019relations}, and self-presentation via Instagram posts and stories~\cite{chiu2021last}.}
However, most of these studies were conducted within a single cultural context or monolingual groups, such as English-speaking communities~\cite{haimson2017makes, ahn2025stan,de2019relations,chiu2021last} or Chinese-speaking communities~\cite{wang2024critical, liu2024wrist,wan2025hashtag,wu2018danmaku}, rather than examining the dynamics of communication across cultural and linguistic boundaries~\cite{bartolome2023literature}.
However, as international platforms increasingly mediate interactions among diverse populations, cross-cultural communication is emerging as a critical yet underexplored topic. The recent influx of ``TikTok refugees'' into RedNote offers a timely opportunity to investigate this gap by examining how Western newcomers and Chinese users directly engage with one another in a shared digital environment.

Among the highly diverse and complex content sparked by the ``TikTok refugee'' event, we chose to focus on one specific topic, namely posts in which foreign newcomers ask Chinese users to give Chinese names (\#Give me a Chinese Name\#).
\textcolor{black}{We chose this topic because naming is often the first step of starting a dialogue, yet it remains a particularly intricate challenge in cross-cultural communication. Unlike simple labeling, naming functions as a dense social act in which identity marking~\cite{gatson2011self}, performative power~\cite{butler2021excitable}, and symbolic meaning~\cite{alford1987naming} converge (see \autoref{ssec:naming} for more related work). While machine translation can handle surface-level language, it often fails to capture precisely these layered social and cultural dimensions~\cite{bourdieu1991language}. 
This complexity is especially pronounced in the Chinese context due to fundamental differences in naming paradigms: while English naming practices are primarily sound-based, Chinese naming requires selecting characters that both match pronunciation and convey symbolic or aesthetic connotations. This process can yield many potential Chinese names from a single foreign name~\cite{li2007semantic, kim2020linguistics}.
These untranslatable layers of meaning collectively construct a modern ``Babel Tower'', where Chinese users employ complex ``information encoding'' methods that complicate transparent decoding across languages and cultures.
}

To address this issue, we posed the following questions: \textbf{(RQ1)} What information encoding strategies are employed in the naming practice? \textbf{(RQ2)} How are these encoding strategies combined or layered, thereby increasing the cost of cross-cultural communication? \textbf{(RQ3)} What are the frequency and distribution of these strategies?
\textcolor{black}{However, addressing our RQs requires overcoming two major challenges. First, prior work in cross-cultural communication has typically relied on small-scale qualitative or survey-based studies \cite{choi2005qualitative,he2010understanding}, leaving few methodological resources for systematically analyzing large, noisy, multilingual, and high-context corpora. Second, naming practices are densely layered with cultural, semantic, and other contextual cues that are difficult to decode using standard computational methods or translation tools \cite{naveen2024overview}.}

To meet these challenges, we developed a human-in-the-loop approach that leverages the strengths of large language models (LLMs) while retaining human judgment. \textcolor{black}{This approach allowed us to clean, process, and structure 70,614 high-context naming comments that would be infeasible to handle manually. Building on this structured corpus,}
we derived a systematic framework that identifies more than 40 fine-grained naming strategies across semantic, phonetic, and visual channels. We also analyzed how strategies across these channels combine and layer within naming practices, 
\textcolor{black}{as well as their correlation with engagement metrics such as the number of likes. } 
Substantively, our analysis of the ``Babel Tower'' effect offers design implications for translation, recommendation, and governance mechanisms on social media platforms that aim to support multicultural communities. 
\textcolor{black}{To sum up, the key contributions of this paper are:}
\begin{itemize}

\item {\textcolor{black}{\textbf{A human–AI approach for structuring cross-cultural corpora at scale.} We introduce a human-in-the-loop approach that combines LLM crowdsourcing with consistency-guided correction, addressing the HCI challenge of limited data scale while enabling reliable extraction and structuring of 70,614 multilingual, culturally dense naming comments.}}

\item {\textcolor{black}{\textbf{A multi-channel framework and analysis of naming practices in cross-cultural communication.} Using the structured corpus, we develop a systematic framework of semantic, phonetic, and visual naming strategies, examine how these strategies are combined and the engagement patterns of Chinese users, and use this ``Babel Tower'' as empirical leverage to extend and critically reflect on traditional cross-cultural theories.}}

\item {\textcolor{black}{\textbf{Design and methodological implications for multicultural platforms.} Our findings demonstrate, at the research level, that the proposed human–AI approach can generalize to other cross-cultural datasets, and, at the practical level, that it can inform design recommendations for platform translation and governance mechanisms that better support multicultural communities.}}


\end{itemize}

\textcolor{black}{Before proceeding, we note that the corpus analyzed in this paper consists of user-generated images, posts and comments from RedNote and may contain slurs, profanity, or other potentially offensive expressions. The presence of these examples does not reflect the authors' views and should not be interpreted as endorsing any of the attitudes or positions expressed by platform users.}
\label{sec-1}

\section{ Related Work}\label{sec-2}
This section reviews three important areas of research that inform our work.

\subsection{Social Media Research in HCI}
With the rise of social media platforms such as Facebook, Twitter, Instagram, and TikTok, these apps and interfaces have become important sites of human-computer interaction. In the field of HCI, research on social media is often conducted within the framework of computer-mediated communication (CMC)   ~\cite{lipschultz2022introduction}. Previous studies have examined how digital platforms mediate interaction in various ways, including text-based sociality (e.g. comments and posts)~\cite{lampe2004slash}, interactive practices (e.g., instant messaging, likes, bullet comments)~\textcolor{black}{\cite{nardi2000interaction,jiang2024games}}, and short video platforms such as TikTok and YouTube, where interaction is shaped by video affordances and algorithmic recommendation~\cite{bartolome2023literature}. This work has established CMC as a key lens for understanding how technologies structure communication and meaning-making in digital environments.
More recently, HCI research has extended these concerns to social media contexts, where communication unfolds through large-scale and synchronous forms of interaction such as comment streams, live stream chats and bullet screen ``danmu'' messages \textcolor{black}{\cite{hamilton2014streaming,wu2018danmaku}}. Studies show how such practices shape visibility, foster communal engagement, and regulate participation in highly dynamic digital spaces \textcolor{black}{\cite{seering2017shaping,lu2018you}}. For example, research on danmu highlights how rapid on-screen commentary creates both shared affect and novel coordination challenges in online communities~\cite{wu2018danmaku}.

Despite these advances, much of the literature remains bounded within single-language or single-cultural contexts. Analyses of platforms such as Twitter, Reddit, and YouTube often focus exclusively on English-speaking users, while studies of Weibo or Bilibili attend primarily to Chinese contexts \textcolor{black}{\cite{zhang2020making,tufekci2014big,lu2018you}}. As a result, cross-cultural and multilingual dynamics, in which diverse linguistic repertoires intersect and meanings are continuously negotiated, have received relatively limited attention in HCI research.
A particularly tricky communication scenario arises around user-generated translation and naming practices. These studies highlight how social media participants actively produce situated meanings through playful translation, transliteration, and renaming strategies~\cite{lu2021understanding,leppanen2014entextualization}. Among these practices, naming stands out as a particularly salient cultural act: names operate as symbols of identity and power, and in cross-cultural contexts they often require negotiation across linguistic boundaries \cite{alia2007names,van2017usernames,girma2020black}.

While cross-cultural dynamics remain underexplored, the naming practices that emerged during the ``TikTok refugee'' event provide a powerful lens for examining how communication unfolds across linguistic and cultural boundaries. In this work, we use naming strategies to investigate the construction of a ``Babel Tower'' in cross-cultural communication.

\subsection{
\textcolor{black}{Naming Practice in} Cross-Cultural Communication}
\label{ssec:naming}
\textcolor{black}{Naming uniquely concentrates several key dimensions of cross-cultural communication within a single practice. As identity markers, names are social acts through which people express authenticity, signal cultural background, and position themselves within specific communities~\cite{gatson2011self}. In cross-cultural settings, these identity cues depend on culturally specific associations that are not encoded in surface language~\cite{alford1987naming}. Naming is also a performative act that assigns social positions rather than merely labeling individuals~\cite{butler2021excitable,foucault2013archaeology}: the act of giving a name can include or exclude others, mark them as insiders or outsiders, and place them within classificatory systems that define group boundaries and visibility. These power effects rely on culture-specific categories and implicit norms that machine translation does not model. In addition, personal names pose distinctive transliteration challenges, since they must balance phonetic similarity with structural constraints and culturally salient meanings~\cite{karimi2011machine,li2004joint}. Whereas English naming practices are primarily sound-based, Chinese naming requires choosing characters that fit pronunciation while also conveying symbolic or aesthetic connotations, so a single foreign name can yield multiple Chinese forms with different implied identities~\cite{li2007semantic,kim2020linguistics}. Finally, names carry symbolic connotations related to personality, traits, and social identity: sound patterns and character choices are interpreted as cues to temperament, status, and group membership, making names culturally meaningful symbols rather than neutral labels~\cite{sidhu2015s,goldstein2016patrick,blum1997naming}.
}
Edward T. Hall proposed the ``high-context/low-context cultural theory'', which reveals how different cultural backgrounds influence communication patterns. According to Hall's framework, cross-culture naming practices is typically categorized as high-context, where meanings are inferred from implicit cues and shared understandings {\cite{hall1976beyond}}.
\textcolor{black}{These indicate that cross-cultural naming both reflects underlying cultural differences and exposes the challenges of mutual understanding.} 

So far, prior HCI research on cross-cultural interactions primarily focuses on computing technology applied to facilitate collaboration among multi-cultural team members \cite{kim2002cross,diamant2008did,aiken2002multilingual}, as well as the role of machine translation in supporting multilingual interaction \cite{zhang2022facilitating,gao2014beliefs,wang2013machine}. In addition, some studies exploring how interaction design adapts to cross-cultural differences \cite{baughan2021cross} and developing practical tools for better communication \cite{baughan2021cross,xu2014improving,rodolico2022understanding}. 

However, most prior studies have treated cross-cultural factors as background conditions, focusing on efficient information transmission rather than interpreting cultural meaning or examining deeper interactional processes. \textcolor{black}{Given the significance of naming in cross-cultural interaction,} our work addresses this gap by analyzing naming practices among Eastern and Western users, identifying multi-channel naming strategies and how their combinations produce a ``Babel Tower'' effect. By further linking these strategies to user engagement patterns, we show how naming serves as a microcosm of broader cross-cultural communication dynamics.

\begin{figure*}[t!]
    \centering
    \includegraphics[width=\linewidth]{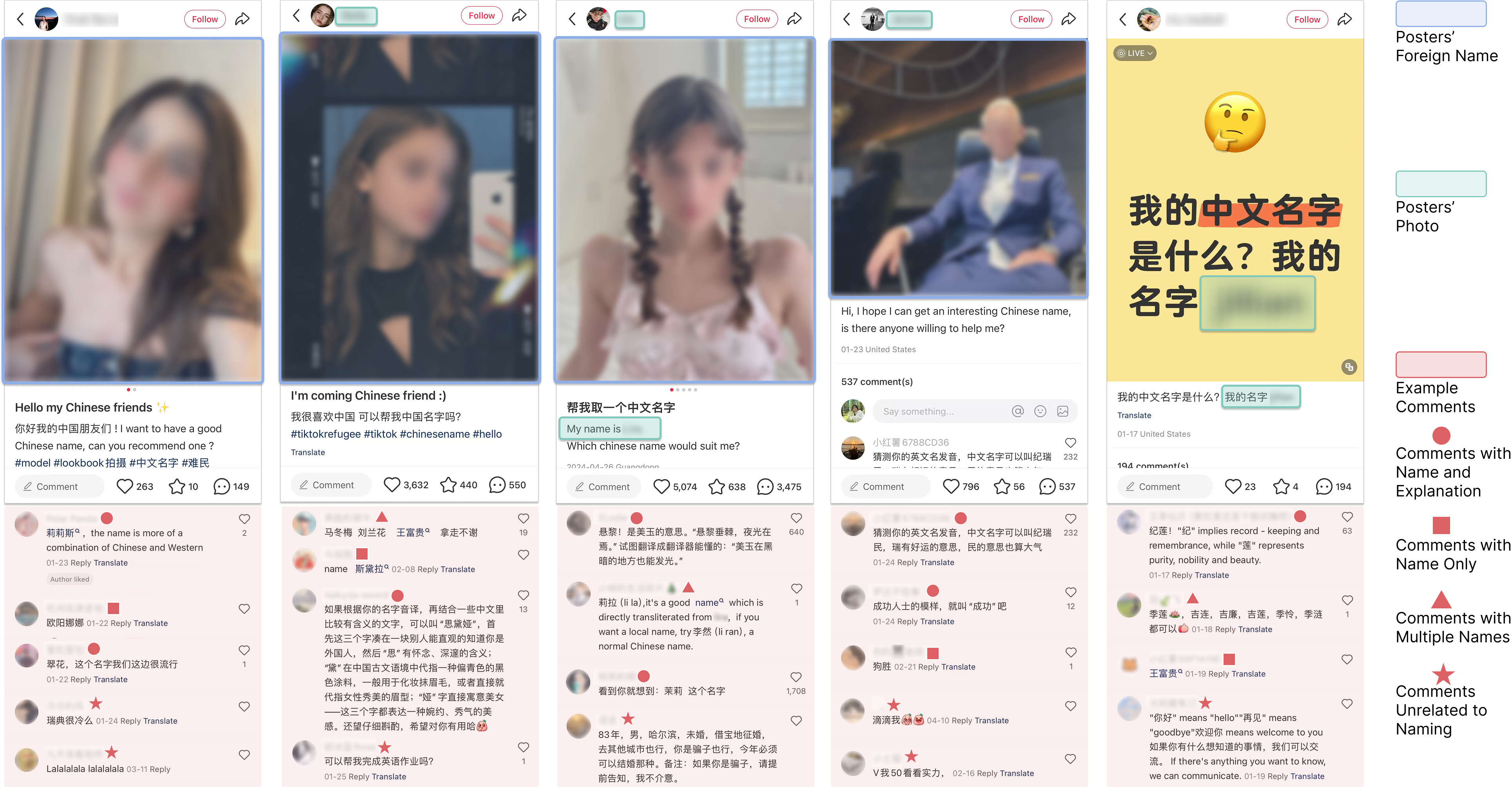}
    \caption{
    Example posts and comments collected from the naming practice. Blue boxes indicate the posters' photos, green boxes indicate the posters' foreign names, and red boxes indicate example comments associated with each post.
    }
    \label{fig:Data}
    \Description{Example posts and comments collected from the naming practice. Blue boxes indicate the posters’ photos, green boxes indicate the posters’ foreign names, and red boxes indicate example comments associated with each post.}
\end{figure*}

\subsection{LLMs for Content Analysis}
LLMs have become central to text analysis, demonstrating strong performance in information extraction, generation, and annotation. Xu et al. \cite{xu2024large} survey generative approaches to relation and event extraction, while Tan et al. \cite{tan2024large} emphasize their role in corpus augmentation and annotation. He et al. \cite{he2023annollm} show that prompting strategies can reach annotation quality comparable to human workers, and Kasner et al. \cite{kasner2025large} establish LLMs as cost-effective and consistent annotators. Collectively, these works demonstrate how LLMs are reshaping methodological foundations in text analysis. Despite these advances, limitations emerge in cross-cultural contexts. Zhang et al. \cite{zhang2025speculating} reveal ``data contamination'' in Chinese corpora, while Yang et al. \cite{yang2025cultural} show that divergences in metaphors and symbolic expressions exacerbate interpretive bias. To address these issues, Xiao et al. \cite{xiao2025enhancing} propose retrieval-augmented reasoning for Chinese social media, enhancing robustness in handling non-standard expressions. Together, these studies highlight the challenges of applying LLMs in heterogeneous cultural settings.

To mitigate bias and enhance performance, two main strategies have been proposed: model self-evaluation and reinforcement, and human-AI collaboration \cite{carnat2024human}. The former, exemplified by recent work on instruction-tuning and self-critique \cite{tseng2025evaluating,ho2025self,gou2023critic,zhou2025refinecoder}, shows potential in expert domains, but its consistency remains limited. Consequently, more attention has turned to human-AI collaboration. Pangakis and Wolken \cite{pangakis2024keeping} highlight human-centered annotation to reduce bias, Wang et al. \cite{wang2024human} design verification workflows that integrate human checks, and Choksi et al. \cite{choksi2024under} demonstrate GPT-4's value in qualitative coding. These studies illustrate the promise of collaborative pipelines, yet they also reveal gaps: current approaches seldom define where human input should be embedded within workflows or how to calibrate confidence across stages, underscoring the need for more systematic frameworks. 

Our work extends this body of literature by addressing a unique form of cross-cultural communication data that embed rich cultural metaphors and non-standard semantic meanings. To this end, we propose a creative human-in-the-loop approach that balances efficiency and reliability in cross-cultural annotation.


\section{Data}
\label{sec-3}

Our study centers on a unique interactional practice on RedNote, where foreign newcomers request Chinese users to give them Chinese names. 
The dataset was collected within days of the event, capturing the earliest stage of interaction before deletions, account closures, or platform restrictions, thereby preserving a snapshot of this moment that would be difficult to reconstruct retrospectively.

To construct our dataset, we adopted a tag-based search strategy, \textcolor{black}{a common and validated method in HCI research for collecting social media content~\cite{devito2017algorithms,chancellor2016thyghgapp, wan2025hashtag}. We used both English and Chinese keywords to ensure coverage and diversity,} searching for tags such as
\textit{\#Give me a Chinese Name\#}, 
\textit{\#Chinese Name\#}, 
\textit{\#Can I have a Chinese Name\#}, 
\textit{\#\textnormal{中文名字}\#}, 
and \textit{\#\textnormal{在线征名}\#}, \textcolor{black}{including minor spelling and capitalization variants in both languages. Data collection spanned from January~21 to February~24,~2025.} All retrieved posts were manually verified by two authors.
Posts were retained only if their content explicitly featured Chinese users giving Chinese names to foreigners. To ensure that the posts reflected active participation rather than isolated cases, we included posts that met at least one of the following criteria: having more than 200 likes, more than 50 comments, or direct interaction from the original poster in the comment section. To avoid duplicates, we collected all eligible posts and performed a deduplication process, ensuring that every post in our dataset was unique.  

As a result, we retained 318 valid posts and 70,614 associated comments for analysis.
The dataset was obtained legally and contains no real-identity information about users \textcolor{black}{(reviewed and approved by the IRB in our school). Note that due to platform API restrictions, only first-level comments (i.e., direct replies to the post) were accessible, while nested replies cannot be collected. More importantly, from a research standpoint, nested replies present a fundamental ambiguity: the intended target of a response—whether it is the original post, a first-level comment, or another nested reply—is often impossible to determine with certainty. Given this combination of technical inaccessibility and interpretive ambiguity, our analysis was finally focused on direct responses to the posts.}


Because the information provided by the post authors may influence Chinese users' naming strategies, we supplemented the dataset by scraping another two types of information: the poster's portrait photo (if provided) and self-reported foreign name (if provided). \textcolor{black}{We added an anonymization step for the poster's portrait photo to obscure key facial features (e.g., eyes, mouth).} To ensure accuracy, one author extracted this information and another author cross-checked the results. For each post, we examined the post content, the profile avatar, and the personal homepage. We defined a ``portrait photo'' as a human face photograph, excluding icons, logos, cartoons, scenery, or brand marks. A ``foreign name'' was recorded only if it was explicitly self-reported in the post or profile, when the information was ambiguous, the field was left blank. For example, in the third post of ~\textcolor{black}{\autoref{fig:Data}}, the poster uploaded a photo with the text ``My name is Lira. Which Chinese name suit me?''. We extracted the photo, documented as the poster's photo, and recorded ``Lira'' as the poster's foreign name. In our final dataset, 193 posts contained both a photo and a foreign name, 58 lacked the photo, 19 lacked the foreign name, and 48 lacked both. 

To sum up, for the 318 posts we collected, we documented three fields: \textit{post link}, \textit{poster's photo}, and \textit{poster's foreign name}; for the 70,614 associated comments with these posts, we documented: \textit{comment content} and \textit{number of likes}.

\section{Method}
\label{sec-4}

\begin{figure*}[t]
    \centering
    \includegraphics[width=\linewidth]{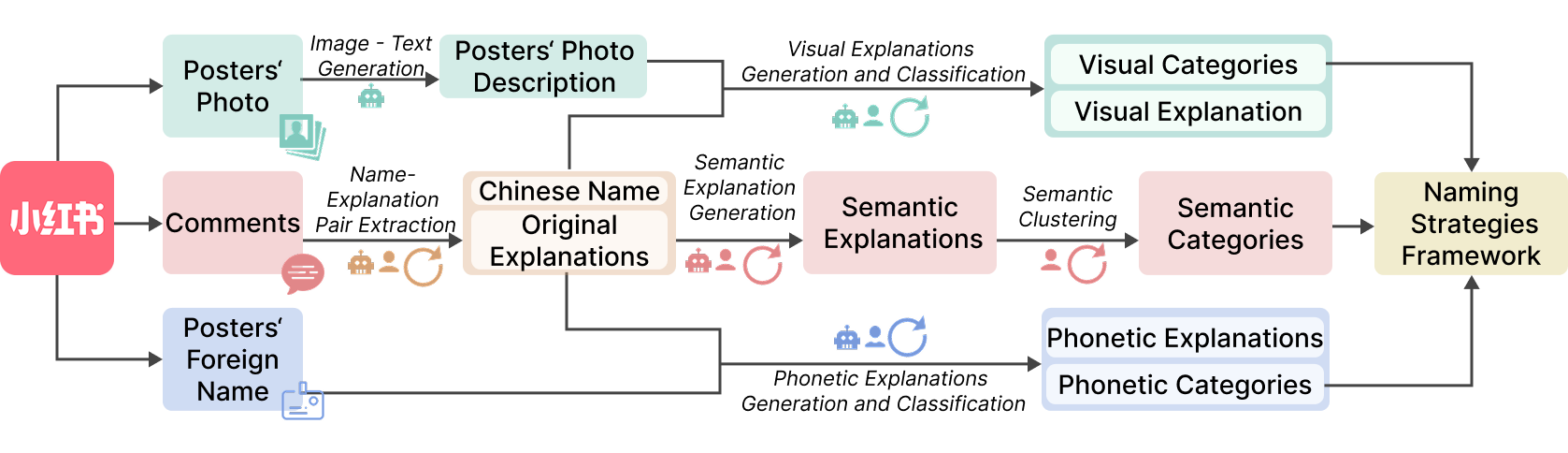}
    \caption{
    \textcolor{black}{The process from obtaining comment data on RedNote to constructing naming strategies framework.}
    }
    \label{fig:Pipeline}
    \Description{The process from obtaining comment data on RedNote to constructing naming strategies Framework.}
\end{figure*}

\textcolor{black}{To analyze the large, noisy, and multilingual dataset, in the following, we describe how we extracted meaningful information from the raw data using a human-AI collaborative approach.} 
The overall pipeline is illustrated in \autoref{fig:Pipeline}.

\subsection{Crowdsourcing LLMs with Consistency-Guided Correction: A Human-in-the-Loop Approach for Extracting Name–Explanation Pairs}
\label{sec-4.1}
Extracting name-explanation pairs is a necessary prerequisite for analyzing the encoding strategies in Chinese naming practices.
Given the scale of over 70,000 multilingual and noisy comments (as shown in \textcolor{black}{~\autoref{fig:Data}}), 
purely manual coding would be prohibitively costly, so we randomly sampled 5,000 comments as a development set to explore AI-assisted extraction. 
Drawing on prior work in processing datasets through human-AI collaboration~\cite{carnat2024human}, and in particular on human-centered coding processes~\cite{pangakis2024keeping, wang2024human, choksi2024under}, we conducted six rounds of iterative refinement and subsequently designed a human-in-the-loop approach. 

\subsubsection{Iterative Refinement of the Extraction Approach}
We conducted six rounds of iterative refinement. The following parts describe each round in detail. 

\textbf{Initial Manual Attempt and 5,000-Comment Development Set for AI-Assisted Extraction.}  
We initially attempted to extract name-explanation pairs manually from comments. 
Two coders independently annotated the same subset of comments to identify naming-related content and record both the proposed name and its explanation. Conflicts were resolved through discussion to reach a consensus.
Each coding took about 40 seconds on average, making manual extraction from over 70,000 multilingual and noisy comments extremely labor-intensive. Moreover, this workload could lead to coder fatigue, reducing coding accuracy. 
These challenges made full manual extraction unfeasible and motivated us to explore AI-assisted extraction.
For experimentation, we randomly sampled 5,000 comments from the full dataset to construct a development set. 
\textcolor{black}{The representativeness of the 5,000-comment sample was confirmed against the full corpus (N = 70,614) by comparison of post time and number of likes (Kolmogorov-Smirnov tests, $p = .171$ and $p = .680$, respectively), as well as the proportion of Chinese versus other languages (chi-square test, $p = .166$).
All p-values exceeded 0.05, suggesting that the subset reasonably reflects the temporal, linguistic, and engagement patterns of the full dataset (Sec. S6 in the supplementary material).}
This set was then used to iteratively refine our approach, as described below.

\textbf{Iteration Round 1: Baseline.}
Given that our dataset primarily consists of Chinese comments containing cultural references, idioms, and slang, we selected DeepSeek-V3 as the baseline model. DeepSeek-V3 has been shown to excel in Chinese-specific tasks, as it assigns more training tokens to learn Chinese knowledge, resulting in better performance on tasks that require Chinese cultural and linguistic understanding \cite{liu2024deepseek}. 
At this stage, our understanding of the scope of the comments was still limited, so we designed a simple prompt instructing the model to extract any person \textit{names} and their associated \textit{explanations}. If no explanations was present, the explanation field was set to null; if neither a name nor a description appeared, both fields were set to null.
\textcolor{black}{Two authors, both thoroughly familiar with the task and dataset, independently reviewed each name–explanation pair generated by DeepSeek-V3. For each pair, they examined the model-extracted fields and recorded their own corrected version in a separate column if they disagreed with the model output. 
We first assessed inter-coder reliability on these independent judgments, finding substantial agreement~(Cohen's κ = 0.67) \cite{landis1977measurement}.
The two sets of judgments were then compared to determine the final labels: if both authors accepted the model output, it was adopted as the final label (agreement to accept); if both rejected it but proposed the same correction, that version was used (agreement to correct); cases were flagged as disagreements if the coders diverged—either one accepted the model output while the other proposed a correction, or they proposed different corrections.
All disagreements were resolved through a joint review of the original comment and a discussion of rationales until a consensus was reached on a final version. This consolidated set of labels forms our gold-standard dataset.} 
\textcolor{black}{During reconciliation, we categorized each corrected case by error type, and these correction patterns~\autoref{tab:errors} guided both our final consensus labels and subsequent iterations.}
The accuracy of DeepSeek-V3 in this initial round, measured against the gold standard, was 67.84\%.

\begin{table*}[t]
\centering
\renewcommand{\arraystretch}{1}
\caption{Error patterns, examples, and correction rules for the name--explanation extraction task.}
\label{tab:errors}

\begin{tabular}{
>{\raggedright\arraybackslash}p{0.6cm}
>{\raggedright\arraybackslash}p{5.2cm}
>{\raggedright\arraybackslash}p{3.4cm}
>{\raggedright\arraybackslash}p{5.4cm}
}
\toprule
\textbf{ID} & \textbf{Error Patterns} & \textbf{Example} & \textbf{Correction Rule} \\
\midrule
1 & Explanations contain emojis or emoticons that confuse interpretation
  & ``好厉害 [emoji][emoji]''
  & Remove emojis or emoticons from the explanation \\
\midrule
2 & User @-mentions are extracted as names
  & ``@张三 你真棒''
  & Set name and explanation to \textit{null} \\
\midrule
3 & Single-character entries, which are invalid in Chinese naming (surname + given name)
  & ``晴'', ``马''
  & Set name and explanation to \textit{null} \\
\midrule
4 & Purely foreign names, which are outside the scope of Chinese naming practices
  & ``John''
  & Set name and explanation to \textit{null} \\
\midrule
5 & Non-assigned names mentioned in explanations
  & ``张华，把李华位置抢了''
  & Keep ``张华'' as the name; set ``李华'' to \textit{null} \\
\midrule
6 & Redundant or nested extractions
  & ``张谙艺'' and ``谙艺''
  & Keep the full name only; set redundant entry to \textit{null} \\
\midrule
7 & Explanations that are meaningless or unrelated
  & ``马赫 比较好一些''
  & Set explanation to \textit{null} \\
\midrule
8 & Unusual or playful names not recognized by the model
  & ``萝卜土豆切吧切吧''
  & Manually correct into a valid name and explanation \\
\bottomrule
\end{tabular}

\Description{Error patterns, examples, and correction rules for the name--explanation extraction task.}
\end{table*}

\textbf{Iteration Round 2: Data processing + Prompt enhancement.}
From the manual verification in Round 1, we identified recurring error patterns and summarized eight error types to guide subsequent refinements (~\autoref{tab:errors}).
Based on these patterns, we made improvements through three steps. \textit{Preprocessing} removed emojis, emoticons, and user @-mentions in each explanation, addressing \textbf{\textit{Error 1}} and \textbf{\textit{Error 2}}. \textit{Prompt enhancement} added explicit background that the task involves Chinese users on RedNote giving Chinese names to foreigners, highlighting the multicultural context and the restriction to Chinese names. This addressed \textbf{\textit{Error 4}} and supported \textbf{\textit{Error 8}} by encouraging recognition of unconventional names. We also embedded explicit rules with positive and negative examples, as shown in ~\autoref{tab:errors}, which addressed \textbf{\textit{Errors 3–8}}. \textit{Postprocessing} further checked the outputs, and any remaining single-character or foreign names were set to null, addressing \textbf{\textit{Error3}} and \textbf{\textit{Error 4}}.
We then ran the extraction with DeepSeek-V3 and evaluated its accuracy against the gold standard. Outputs identical to the gold standard were accepted as correct directly. The remaining cases were manually reviewed. 
\textcolor{black}{During the review, we defined outputs that differed only in surface form but preserved the same semantic meaning (e.g., variations in punctuation, quotation marks, spacing, or synonymous phrasing) as minor differences and accepted them as correct. For example, ```景明'取自`春和景明'的意境'' and ```景明'是取自`春和景明'的意境'' are treated as minor differences but are semantically equivalent, since the additional word ``是'' does not change the meaning of the explanation.} As a result, in this round, 89.08\% of outputs matched the gold standard, an additional 3.08\% were accepted as correct despite minor differences, and the overall accuracy reached 92.16\%. This shows a clear improvement over Round 1 and confirms the effectiveness of data processing and prompt enhancement.

\textbf{Iteration Round 3: Data processing + Prompt enhancement + Crowdsourcing with five independent DeepSeek models.}
Building on the results of Round 1 and 2, we found that even with DeepSeek-V3, some codings varied across iterations. This observation is similar to the behavior of human coders, where individual judgments may differ. 
Motivated by this, we explored whether multiple LLMs could simulate the process of crowdsourced human coding. Prior studies have shown that using multiple LLMs can expand coding capacity and reduce variance in final labels~\cite{ziems2024can,tian2024theory}, but no existing work has applied this strategy in a culturally specific context such as Chinese naming practices.  
Building on Round 2, we continued to use the same data processing and enhanced prompt design. To simulate human coding, where consensus is usually reached through majority voting, we employed five independent DeepSeek-V3 models to annotate each comment. The choice of five models was informed by findings from linguistic and culturally aware coding studies, which reported that each item is typically coded between two and ten times \cite{huang2023culturally, chakrabarty2023spy, mao2024metapro, joseph2023newsmet, dou2022improving}. Selecting five offered a balance between diversity of outputs and computational feasibility. All models were run with the same prompt and settings, reflecting the assumption that human coders receive the same training and share a common understanding of the task.  
The outputs for each item were aggregated through majority voting. Let $o_{1}, o_{2}, \dots, o_{5}$ denote the results from the five models. The final label $\hat{o}$ is defined as  
\[
\hat{o} = \text{argmax}_{o \in \{o_1, \dots, o_5\}} \; \text{count}(o),
\]
where $\text{count}(o)$ represents the number of times output $o$ appears. When the maximum count was not unique, the final label was selected uniformly from the tied candidates.  
With this scheme, 88.66\% of outputs matched the gold standard exactly, and an additional 3.62\% were accepted as correct despite minor differences, yielding an overall accuracy of 92.28\%. Compared with Round 2, the performance gain was small, suggesting that multiple DeepSeek models added limited performance benefit, though they increased our confidence in the results' consistency.  

\begin{table*}[t]
\centering
\caption{Distribution of consistency scores across coding outcomes for identical vs different models ($\chi^2$ test, $p < .01$).}
\label{tab:consistency_comparison}

\begin{tabular}{lccccc}
\toprule
\multicolumn{6}{l}{\textbf{(a) Consistency scores across five independent DeepSeek models}}\\
\midrule
Outcome & 0 & 40 & 60 & 80 & 100 \\
\midrule
Correct (Matched gold) & 0 (0\%) & 3 (0.07\%) & 189 (4.26\%) & 164 (3.70\%) & 4077 (91.97\%) \\
Correct (Minor diff.)  & 0 (0\%) & 6 (3.31\%) & 28 (15.47\%) & 21 (11.60\%) & 126 (69.61\%) \\
Incorrect              & 0 (0\%) & 2 (0.52\%) & 55 (14.25\%) & 46 (11.92\%) & 283 (73.32\%) \\
\midrule
\multicolumn{6}{l}{\textbf{(b) Consistency scores across five different and independent LLMs}}\\
\midrule
Outcome & 0 & 40 & 60 & 80 & 100 \\
\midrule
Correct (Matched gold) & 2 (0.04\%) & 55 (1.22\%) & 230 (5.11\%) & 385 (8.55\%) & 3833 (85.08\%) \\
Correct (Minor diff.)  & 10 (3.46\%) & 89 (30.80\%) & 105 (36.33\%) & 54 (18.69\%) & 31 (10.73\%) \\
Incorrect              & 12 (5.83\%) & 58 (28.16\%) & 80 (38.83\%) & 39 (18.93\%) & 17 (8.25\%) \\
\bottomrule
\end{tabular}

\Description{Distribution of consistency scores across annotation outcomes for identical vs different models ($\chi^2$ test, $p<0.01$).}
\end{table*}

\textbf{Iteration Round 4: Data processing + Prompt enhancement + Crowdsourcing with five independent DeepSeek models + Consistency-guided human correction.}
Building on Round 3, we calculated the consistency score of outputs from five DeepSeek-V3 models to quantify agreement. The score was defined as  
\[
\text{Consistency}(o_1, \dots, o_5) = \frac{\max_{o}\, \text{count}(o)}{5} \times 100,
\]
yielding consistency scores of 100 (all identical), 80, 60, 40, or 0. With the crowdsourced DeepSeek-V3 models, 89.72\% of cases had a score of 100, 4.62\% had 80, 5.22\% had 60, 0.22\% had 40, and none had 0. Since the vast majority of items reached full agreement, we accepted cases with a score of 100 directly and conducted human review on those with lower scores.
With this human-AI collaborative scheme, an additional 2.06\% of cases were manually corrected from errors, which raised the overall accuracy from 92.28\% in Round 3 to 94.34\%. ~\autoref{tab:consistency_comparison} (a) further shows that consistency scores were significantly associated with coding correctness (χ² test, $p < .01$). High consistency (100) dominated correct cases (91.97\%), whereas lower scores appeared more often in minor difference and incorrect cases, confirming that lower agreement signals greater task difficulty and ambiguity. Overall, Round 4 demonstrated the effectiveness of such a human-AI collaborative approach.

\textbf{Iteration Round 5: Data processing + Prompt enhancement + Crowdsourcing with five different and independent models.}
Although the AI-human collaborative approach in Round 4 improved accuracy, we found that using five DeepSeek-V3 models still had limited ability to identify difficult and ambiguous cases. This limitation may stem from the fact that all models shared the same architecture and training data, leading to similar judgments on the same task. In contrast, human annotators may receive the same training and share a common understanding of the task, but they also bring different experiences, backgrounds, and domain expertise. We therefore hypothesized that using five different LLMs for crowdsourcing would better reflect the differences of human coders.  
To test this, we selected five recent leading models with different architectures and training data: DeepSeek-V3\footnote{\url{https://arxiv.org/abs/2412.19437}}, GPT-4.1-mini\footnote{\url{https://help.openai.com/en/articles/9624314-model-release-notes}}, Grok-3-mini\footnote{\url{https://x.ai/news/grok-3}}, Gemini 2.0 Flash\footnote{\url{https://storage.googleapis.com/deepmind-media/gemini/gemini_v2_5_report.pdf}}, and Qwen-Plus\footnote{\url{https://arxiv.org/abs/2412.15115}}. Building on the data processing and enhanced prompt design from Round 2, we applied the same crowdsourcing procedure as in Round 4, but replaced the five identical DeepSeek-V3 models with these five different LLMs.  
The results showed that 90.10\% of outputs matched the gold standard exactly, and an additional 5.78\% were accepted as correct despite minor differences, yielding an overall accuracy of 95.88\%. Compared with Round 3, where five identical DeepSeek-V3 models were used, this setting achieved higher performance. These findings suggest that crowdsourcing with heterogeneous LLMs better reflects the variety of perspectives found in human coding and provides a more effective strategy.

\textbf{Iteration Round 6: Data processing + Prompt enhancement + Crowdsourcing with five different and independent models + Consistency-guided human correction.}
Building on Round 5, we further integrated the consistency-guided human correction strategy introduced in Round 4. 
As shown in ~\autoref{tab:consistency_comparison}, crowdsourcing with different models revealed difficult cases more effectively than crowdsourcing with identical models. For minor difference cases, 89.27\% of items could be detected by the consistency score under different models, compared with only 30.39\% under identical models. For incorrect cases, 91.75\% could be detected under different models, compared with 26.68\% under identical models. This shows that different models were less likely to mask errors with unanimous but wrong predictions, and that consistency scores provided a significantly stronger signal for identifying problematic cases. On top of Round 5, an additional 3.78\% of cases were manually corrected from errors, raising the overall accuracy from 95.88\% to 99.66\%. This final iteration therefore combines the benefits of data processing, prompt enhancement, crowdsourcing with different LLMs, and human intervention, and represents the approach we ultimately adopt for extracting name-explanation pairs (Sec. S1 in the supplementary material).

\subsubsection{Final Application to the Full Dataset}
We applied this human-in-the-loop extraction approach to the remaining 65,614 records. Of these, 22,203 did not achieve full agreement across the five different models and were therefore subject to manual review. Four coders, all fully familiar with the coding guidelines, the task, and the dataset, carried out the work. Three coders each verified and corrected 5,500 records, and the fourth processed 5,503 records. In total, 2,535 extraction errors were corrected, accounting for 3.86\% of the dataset. The corrected records were then combined with the previously annotated 5,000 gold-standard records to produce the complete ``name-explanation'' pair dataset. This dataset serves as the foundation for the subsequent tasks.

\begin{table*}[ht]
\centering
\caption{Comparison of accuracy, efficiency, and human effort across iterations. 
Accuracy was evaluated on the 5,000-comment development set, while efficiency was estimated as the time required to process the full dataset of 70,614 comments. 
Definition: Baseline (DeepSeek-V3 with a simple prompt), Data processing (D), Prompt enhancement (P), Crowdsourcing with identical DeepSeek models (CD), Crowdsourcing with different models (CF), Consistency-guided human correction (H).}
\label{tab:iteration_summary}
\begin{tabular}{lccc}
\toprule
\textbf{Iteration} & \textbf{Accuracy (\%)} & \textbf{Efficiency (hours)} & \textbf{Human effort} \\
\midrule
Round 0: Manual extraction & – & Estimated 784.6 & Direct coding by at least 3 coders \\
Round 1: Baseline & 67.84 & 14.1 & None \\
Round 2: D+P & 92.16 & 21.2 & None \\
Round 3: D+P+CD & 92.28 & 21.2 & None \\
Round 4: D+P+CD+H & 94.34 & 21.2 & Manual review of flagged cases \\
Round 5: D+P+CF & 95.88 & 23.6 & None \\
Round 6: D+P+CF+H & 99.66 & 23.6 & Manual review of flagged cases \\
\bottomrule
\end{tabular}
\Description{Comparison of accuracy, efficiency, and human effort across iterations. Accuracy was evaluated on the 5,000-comment development set, while efficiency was estimated as the time required to process the full dataset of 70,614 comments. Definition: Baseline (DeepSeek-V3 with a simple prompt), Data processing (D), Prompt enhancement (P), Crowdsourcing with identical DeepSeek models (CD), Crowdsourcing with different models (CF), Consistency-guided human correction (H).}
\end{table*}

\subsubsection{Trade-off Analysis of Different Approaches}
We analyzed the trade-offs among accuracy, efficiency, and human effort across different approaches, as illustrated in ~\autoref{tab:iteration_summary}. Purely manual coding may achieve high accuracy, but it is prohibitively costly in both time and labor. Given that coding a single record takes about 40 seconds, we estimated that processing the full dataset would require 784.6 hours, and coder fatigue could severely compromise reliability. In contrast, fully automated approaches (Rounds 1, 2, 3, and 5) removed human effort and achieved substantial efficiency, but their accuracy remained limited. Among these, using multiple different models performed better than single model or identical model crowdsourcing.
The most reliable outcomes were obtained through human-AI collaboration (Round 6), where low-consistency cases were selectively reviewed and corrected by human coders. This strategy greatly reduced the human burden compared to full manual coding while ensuring more dependable accuracy. 

\subsubsection{Model Performance Comparison}
In addition, we compared the individual performance of different models through Round 5. For this analysis, we report only exact matches with the gold standard, as identifying minor-difference but correct cases would require extensive manual review across all models, which was beyond the scope of this supplementary evaluation. The results showed that DeepSeek-V3 achieved the highest accuracy (90.24\%), followed by Qwen-Plus (88.78\%), Grok-3-mini (88.04\%), GPT-4.1-mini (86.32\%), and Gemini 2.0 Flash (83.94\%), confirming that DeepSeek-V3 has relative strengths in processing Chinese text and culturally nuanced content. Based on this evidence, we continued to use DeepSeek-V3 as the primary model for subsequent tasks.

\subsection{Channel-Specific Generation for Missing and Incomplete Explanation}
After extracting the name-explanation pairs, we found that most records contained only a name without a corresponding explanation, while some included explanation that were incomplete or lacked sufficient detail. To support subsequent analysis, it was therefore essential to supplement and refine these missing or incomplete explanations. This task is more challenging than pair extraction in \autoref{sec-4.1}. Generating reliable explanations requires integrating linguistic and cultural meanings and expressing them in coherent, context-appropriate language, which makes manual annotation infeasible and motivates our AI-assisted approach. In RedNote's social media context, explanations often extend beyond the name itself, drawing on attributes of the poster (e.g., appearance or original foreign name). To capture this diversity, we structure explanations along three channels: semantic, visual, and phonetic.

\subsubsection{Name-Explanation Pair Processing}
\label{sec-4.2.1}
For data preparation, we converted the raw output of the name–explanation extraction step into a consistent pair-level format. Because a single comment could contain multiple name–explanation pairs, we first split each multi-pair comment into separate pair records. We then removed records where both the name and explanation were empty. After this normalization and cleaning, the 5,000-comment development set resulted in 4,727 valid pairs, and the remaining data yielded 59,364 valid pairs, totaling 64,091 pairs used for explanation generation.

\subsubsection{Explanation Generation across Three Channels}
\label{sec-4.2.2}
The missing or incomplete explanations were generated through inference across three channels, namely semantic, visual, and phonetic.

\begin{table*}[t]
\centering
\renewcommand{\arraystretch}{1}
\caption{Error patterns, examples, and correction rules for the semantic--based explanation generation task.}
\label{tab:name_rules}

\begin{tabular}{
>{\raggedright\arraybackslash}m{4.2cm}
>{\raggedright\arraybackslash}m{5.2cm}
>{\raggedright\arraybackslash}m{5.2cm}
}
\toprule
\textbf{Error Pattern} & \textbf{Example} & \textbf{Correction Rule} \\
\midrule
Subjective judgments
& ``This is a beautiful name''
& Avoid personal opinions; provide only factual or descriptive content \\
\midrule
Vague or generic explanation
& ``This name has cultural meaning''
& Specify concrete sources when referring to cultural references or memes \\
\midrule
Ignoring negative or satirical intent
& Interpreting ``BYD'' only as a car brand
& Remain neutral and acknowledge vulgar or satirical meanings when present \\
\midrule
Factual errors or hallucinations
& Misattributing a meme to the wrong source
& Ground explanations in authentic cultural usage \\
\midrule
Redundant outputs
& ``This name has no special meaning''
& Omit unnecessary or irrelevant explanations \\
\midrule
Overly divergent associations
& Linking a rare word to obscure folklore
& Restrict explanations to widely recognizable associations \\
\midrule
Neglecting the original explanation
& Skipping irony in the commenter's note
& Incorporate the original explanation into the final output \\
\midrule
Misaligned with context
& Treating ``马云'' literally as the entrepreneur
& Infer the most plausible meaning in the social media context \\
\midrule
Forced interpretation
& Claiming ``贝奇'' relates to ``备齐'' without evidence
& Do not generate far-fetched links when no reasonable connection exists \\
\bottomrule
\end{tabular}

\Description{Error patterns, examples, and correction rules for the semantic--based explanation generation task.}
\end{table*}

\textbf{Semantic-Based Explanations.}
We first focused on generating explanations directly from the semantics of names. Each name, along with its original explanation, was provided as input to the DeepSeek-V3 model.
We applied following procedure to the development set. Our initial design used a simple prompt, and the generated outputs were independently evaluated by two coders. Through repeated evaluation and discussion, we identified recurring issues and refined the prompt iteratively. After several rounds, we assigned DeepSeek-V3 the role of a ``name interpreter'' with expertise in Chinese internet culture and naming practices, expressed in a style close to online discourse. The prompt explicitly emphasized the task context: Chinese users on RedNote giving Chinese names to foreigners, and instructed the model to pay attention to phonetic puns, online memes, humorous expressions, and references to celebrities or popular culture. Across the review cycles, we gradually formulated coding guidance in the form of error patterns and correction rules. These are summarized in ~\autoref{tab:name_rules}, which informed the final version of the prompt to generate semantic-based explanations (Sec. S2 in the supplementary material).

\textbf{Visual-Based Explanations.}
We observed that many explanations were visually linked to the poster's photos. 
To incorporate this channel, we collected 212 posts with portrait photo (as described in \autoref{sec-3}). \textcolor{black}{Using deductive coding approach~\cite{fereday2006demonstrating}, we defined eight visual categories (\autoref{tab:visual_rules}) grounded in prior work on visual identity and portraiture \cite{fereday2006demonstrating, jonisova2019portrait, strathern2018portraits}.} Each portrait photo was then converted into a short textual description using GPT-4o-mini, a model with strong visual-to-text reasoning ability. We instructed the model to focus on these eight attributes, and recorded each description as an additional data field called \textit{image description}. 
We then generate the visual explanations. During generation, \textcolor{black}{the Chinese name, its original explanation, and the \textit{image description} were provided as input to DeepSeek-V3.} DeepSeek-V3 was instructed to classify each case into one of eight categories (\autoref{tab:visual_rules}), each with corresponding explanations, following a structured decision process. The model first parses the Chinese name and its explanation to propose candidate visual associations, then verifies whether the Image Description contains clear, explicit supporting cues (subtle or implicit cues are rejected). If multiple cues are present, it selects a single, most salient category and justifies it by citing the exact phrase(s) from the Image Description; if no cue is found, the case is labeled ``no visual association''. Explanations must stay strictly evidence-based, avoid speculative links, and include a self-check against category definitions and the cited visual evidence before finalization (Sec. S3 in the supplementary material). \textcolor{black}{Further details on the iterative refinement process and result quality are provided in \autoref{Sec-4.2.3}.}

\begin{table*}[t]
\centering
\renewcommand{\arraystretch}{1}
\caption{Classification framework for visual-based explanations with definitions and examples.}
\label{tab:visual_rules}

\begin{tabular}{
>{\raggedright\arraybackslash}m{3.2cm}
>{\raggedright\arraybackslash}m{6.2cm}
>{\raggedright\arraybackslash}m{5.4cm}
}
\toprule
\textbf{Category} & \textbf{Definition} & \textbf{Example} \\
\midrule
Demeanor
& The Chinese name is associated with the demeanor or aura suggested by the photo (e.g., calm, fierce).
& \textit{Focused and slightly serious demeanor} $\rightarrow$ ``安睿 (thoughtful) + 肃 (serious)'' \\
\midrule
Facial features
& The name reflects distinctive facial traits (e.g., nose, eyes, face contour).
& \textit{Large round eyes} $\rightarrow$ ``大眼 (Big-Eyed) 丽莎'' \\
\midrule
Hair
& The name references hair style, length, or color.
& \textit{Brown, curly hair} $\rightarrow$ ``卷毛 (Curly Hair)'' \\
\midrule
Skin tone
& The name highlights skin color or tone visible in the photo.
& \textit{Fair skin tone} $\rightarrow$ ``白 (Fair) + 洁 (Pure)'' \\
\midrule
Facial expression
& The name connects to an expression captured in the photo (e.g., smiling, angry, surprised).
& \textit{Smiling expression} $\rightarrow$ ``林乐乐 (Cheerful and smiling)'' \\
\midrule
Clothing / accessories
& The name draws from clothing items or accessories visible in the image.
& \textit{Wearing a dark red hoodie} $\rightarrow$ ``小红帽 (Little Red Hood)'' \\
\midrule
Background
& The name reflects the setting or object in the background of the photo, but only when salient and clearly displayed.
& \textit{Sitting on rocks with mountain scenery in the background} $\rightarrow$ ``明远山 (Distant Mountains)'' \\
\midrule
No visual association
& No relevant visual cue in the photo can support a connection between the name and the image.
& Return \textit{null} \\
\bottomrule
\end{tabular}

\Description{Classification framework for visual-based explanations with definitions and examples.}
\end{table*}

\textbf{Phonetic-Based Explanations.}
To incorporate this channel, we collected 251 posts that explicitly provided foreign names (as described in \autoref{sec-3}). \textcolor{black}{Drawing on prior linguistic and cross-lingual naming literature, we adopted a deductive approach \cite{azungah2018qualitative} and defined four phonetic mechanisms commonly used to relate foreign names to Chinese names, namely full homophony \cite{pilshchikov2016semiotics}, partial homophony \cite{bradlow2010perceptual}, meaning association \cite{pilshchikov2016semiotics}, and no relation. These four categories form the basis of our phonetic scheme (\autoref{tab:phonetic_rules}).}
\textcolor{black}{During generation, the Chinese name, its original explanation, and the foreign name were provided as input to DeepSeek-V3.} DeepSeek-V3 was assigned the role of a phonetics expert familiar with Chinese Pinyin and English phonology, and was instructed to classify each case into one of the four categories,  each with corresponding explanations, following a structured process. The model first converts the Chinese name into Pinyin and compares its syllables with those of the foreign name, allowing reasonable variation across phonetic systems, and assigns a category based on syllabic similarity. If no transliteration relationship is found, it then checks the original explanation for a possible meaning-based association. In all cases, the explanation must be consistent with the analysis, avoid fabricated links, and include a self-check against category definitions and the available evidence before final output (Sec. S4 in the supplementary material). \textcolor{black}{Further details on the iterative prompt refinement and result quality are provided in \autoref{Sec-4.2.3}.}

\begin{table*}[t]
\centering
\renewcommand{\arraystretch}{1}
\caption{Classification framework for phonetic-based explanations with definitions and examples.}
\label{tab:phonetic_rules}

\begin{tabular}{
>{\raggedright\arraybackslash}m{3.2cm}
>{\raggedright\arraybackslash}m{6.2cm}
>{\raggedright\arraybackslash}m{5.4cm}
}
\toprule
\textbf{Category} & \textbf{Definition} & \textbf{Example} \\
\midrule
Full homophony
& The full Pinyin pronunciation of the Chinese name closely matches the foreign name, allowing for minor phonetic shifts such as tone omission or consonant alternation.
& ``Stefan'' $\rightarrow$ ``史德风 (shǐ dé fēng)'' \\
\midrule
Partial homophony
& A meaningful component of the Chinese name (e.g., surname or single character) phonetically matches part of the foreign name.
& ``Maher'' $\rightarrow$ ``马永安 (mǎ Yǒng ān)'' \\
\midrule
Meaning association
& The Chinese name conveys meaning, imagery, or symbolism that aligns with the foreign name, often supported by the original explanation.
& ``Art'' $\rightarrow$ ``张画 (Huà, art paintings)'' \\
\midrule
No relation
& No clear phonetic or semantic connection exists between the Chinese and foreign names.
& Return \textit{null} \\
\bottomrule
\end{tabular}

\Description{Classification framework for phonetic-based explanations with definitions and examples.}
\end{table*}

\subsubsection{Iterative Refinement of Channel-specific Explanations Generation and Final Evaluation}
\label{Sec-4.2.3}
\textcolor{black}{Four coders (C1–C4) participated in the manual review of generated explanations across the semantic, phonetic, and visual channels. C1 and C3 are in the filed of journalism and communication, while C2 and C4 are in linguistics. All four coders are highly fluent in both Chinese and English and were trained on our materials and coding scheme; before formal coding, they completed several pilot trials. C1 and C2 independently evaluated all model outputs based on the criteria listed below. When disagreements occurred, they first sought a local resolution; otherwise, cases were escalated to C3 and C4 until a four-way consensus was reached. This review process conducted three rounds, informed iterative refinement of the prompt strategies and further clarification of category boundaries for each channel.} We adopted three evaluation criteria. The first was \textit{accuracy}, defined as the absence of hallucinations or factual errors, mainly for semantic-based explanations. The second was \textit{consistency with the original post}, used for visual- and phonetic-based explanations, which required a clear and valid link to the poster's attributes. Vague or speculative associations were rated as inconsistent. The third was \textit{contextual appropriateness}, which assessed whether the explanation aligns with the surrounding textual or situational context, making it a contextually plausible interpretation.

In the final iteration round, the evaluation showed that 97\% of explanations met the criterion of \textit{accuracy}, 98.5\% demonstrated \textit{consistency with the original post}, and 96.7\% achieved \textit{contextual appropriateness}. Across all three criteria, 95.2\% of outputs met the full standard. \textcolor{black}{Inter-coder reliability was substantial~\cite{landis1977measurement} for all three dimensions, with Cohen's $\kappa = 0.705$ for \textit{accuracy}, $0.754$ for \textit{consistency with the original post}, and $0.849$ for \textit{contextual appropriateness}, indicating that the prompting strategy and generated explanations were stable and could be applied consistently.} Given the complexity and subjectivity of this task, which exceeded that of name-explanation pair extraction, this performance indicates a good level of reliability.

\subsubsection{Final Application and Insights}
After achieving an accuracy of 95.2\% on the development set, we applied the refined prompts to the remaining 59,364 records, resulting in a total of 64,091 valid generated explanations. The final dataset contains the original comment, the extracted name, the original explanation, and the generated semantic-based, visual-based, and phonetic-based explanations.
\textcolor{black}{In general, this process revealed that interpretive tasks such as generating explanations are more complex and subjective than objective tasks such as entity extraction. Unlike entity extraction, explanation outputs could not be reliably aggregated through multi-model voting because semantic variation was inherently high, especially when interpretations depended on usage context (e.g., as personal names rather than standalone lexical items). Instead, dependable results required iterative prompt refinement in combination with human review to verify contextual plausibility and correct ambiguous cases.}

\section{Results}
\label{sec-5}
\textcolor{black}{With the structured and complete corpus prepared, we next turn to understanding the information encoding strategies in these naming practices across three channels (i.e., semantic, phonetic, visual).} Since the semantic channel is highly complex, we first used clustering to identify semantic naming strategies and then synthesized a framework capturing the classification and distribution across the three channels. We further analyzed how strategies are combined and how Chinese users engage in the naming practice, \textcolor{black}{in order to show the construction of this cross-cultural ``Babel Tower.''}

\subsection{Semantic Pattern Discovery}
\label{sec-5.1}
To uncover the semantic patterns underlying the large volume and diverse range of name explanations, we applied \textcolor{black}{BERTopic~\cite{grootendorst2022bertopic}}, which uses \textcolor{black}{HDBSCAN~\cite{mcinnes2017hdbscan}} as its clustering backend, to produce candidate groups of semantically related items. We tuned two HDBSCAN hyperparameters, \texttt{min\_cluster\_size} and \texttt{min\_samples}, and evaluated configurations using four criteria: the number of clusters, the number of items labeled as noise, the share of the largest cluster, and the silhouette coefficient computed on assigned items. \textcolor{black}{Across a range of plausible parameter settings, we found that the major semantic themes and the higher level category structure remained overall stable, although the exact number and size of low level clusters varied} (Sec. S5 in the supplementary material). Through manual inspection, we found that some configurations with high silhouette scores produced only a small number of clusters, which often conflated distinct patterns within the same group. 
Because of this, clustering results were not treated as definitive classifications but rather as a recall-oriented step that gathered potential patterns for later manual interpretation. Using a configuration with \texttt{min\_cluster\_size=30} and \texttt{min\_samples=3}, we favored finer granularity and reduced the risk of conflating distinct patterns, producing 449 clusters and 10,095 items labeled as noise (denoted as label negative one in BERTopic).

\begin{figure}[htbp]
    \centering
    \includegraphics[width=\linewidth]{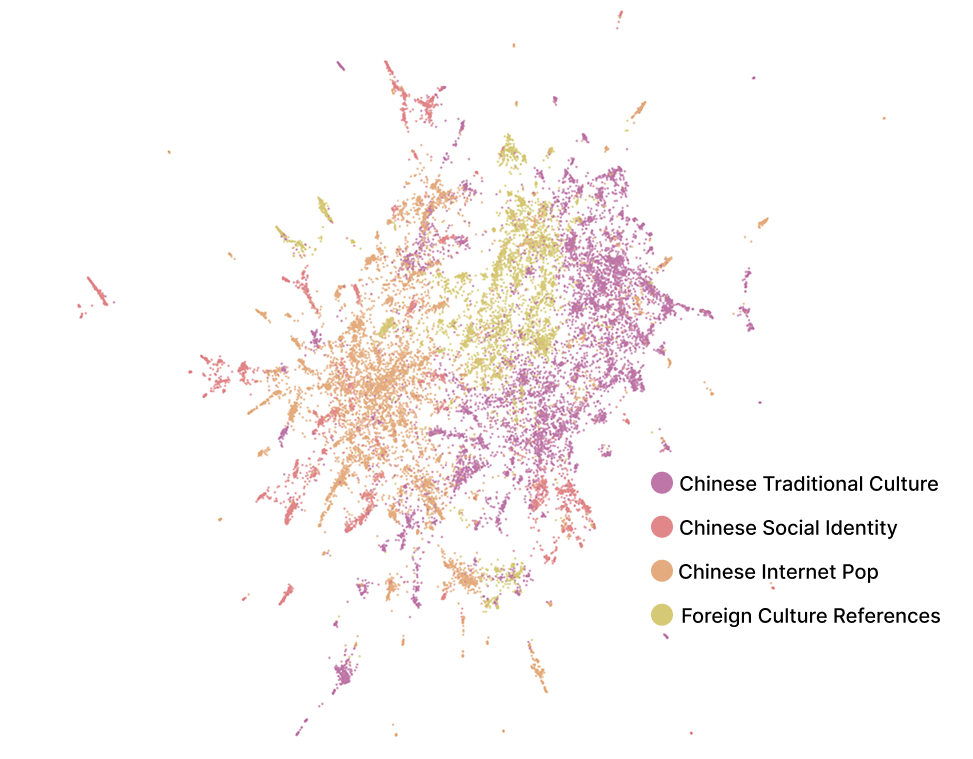}
    \caption{\textcolor{black}{Projection of semantic patterns using BERTopic embeddings and applying UMAP for dimensionality reduction.}}
    \label{fig:Cluster2}
    \Description{2D projection of semantic strategies using BERTopic embeddings for visualization.}
\end{figure}

From these first-stage clusters, four coders iteratively examined and consolidated candidate groups, inductively deriving a framework of 31 semantic naming patterns. 
Because the noise set still contained potentially meaningful patterns, we subjected the 10,095 noise items to a second clustering. We performed a similar parameter sweep and identified \texttt{min\_cluster\_size=5} and \texttt{min\_samples=1} as the most balanced setting for this subset, which yielded 388 clusters (Sec. S5 in the supplementary material). Each of these clusters was inspected against the previously derived 31 categories. Items that aligned semantically with an existing category were integrated, while those that did not fit were explicitly excluded as outliers, \textcolor{black}{\autoref{fig:Cluster2}} shows the 2D projection of the identified patterns,
obtained from BERTopic embeddings \textcolor{black}{and reduced to two dimensions using UMAP \cite{mcinnes2018umap}.} 
 \textcolor{black}{It should be noted that overlaps in the 2D projection do not indicate errors in clustering but reflect semantic and cultural nuances in the data.} Some names are mainly assigned to one category, yet also share meanings or associations with others, so their points appear close to or between clusters. \textcolor{black}{To maintain mutual exclusivity and analytical rigor, coders followed the rules and definitions of the 31 categories in the codebook (Sec. S7 in the supplementary material) to identify each name's primary feature and code it accordingly.} For example, ``翠花'' and ``隔壁老王'' both contain humorous or teasing tones. \textcolor{black}{However, the humor of ``翠花'' draws mainly on folk naming conventions and rural figures, so it is coded as Rural Culture, whereas the humor of ``隔壁老王'' draws on context-free internet jokes and memes, so it is coded as Miscellaneous Memes.}
After final inspection, 895 items remained as residual noise and 735 items were judged unclassifiable, leaving 1,630 outliers in total. The resulting dataset contained 62,461 items mapped to one of the 31 categories, which form the framework of naming strategies along the semantic channel.

\textcolor{black}{To assess the reliability of the 31 semantic naming categories, we conducted a small external expert validation. We randomly sampled 3,000 instances (≈5\% of the final dataset) \cite{gebreegziabher2025supporting}. 
Two external experts participated. One has a background in journalism and communication, and the other is an expert on social media and internet pop culture. Both experts are fluent in Chinese, familiar with Chinese culture, and daily users of RedNote. 
For each sampled instance, we provided the extracted Chinese name, the generated semantic explanation, and its assigned naming category. We also provided the experts with the codebook we developed (Sec. S7 in the supplementary material), which outlines the definitions and decision rules for all categories. 
Each expert independently judged whether the assigned category was correct \cite{ramirez2012topic}. The final evaluation result (overall accuracy = 93.8\%, Cohen's κ = 0.77) suggesting that the category assignments have reasonably good reliability.}

\begin{figure*}[htbp]
    \centering
    \includegraphics[width=\linewidth]{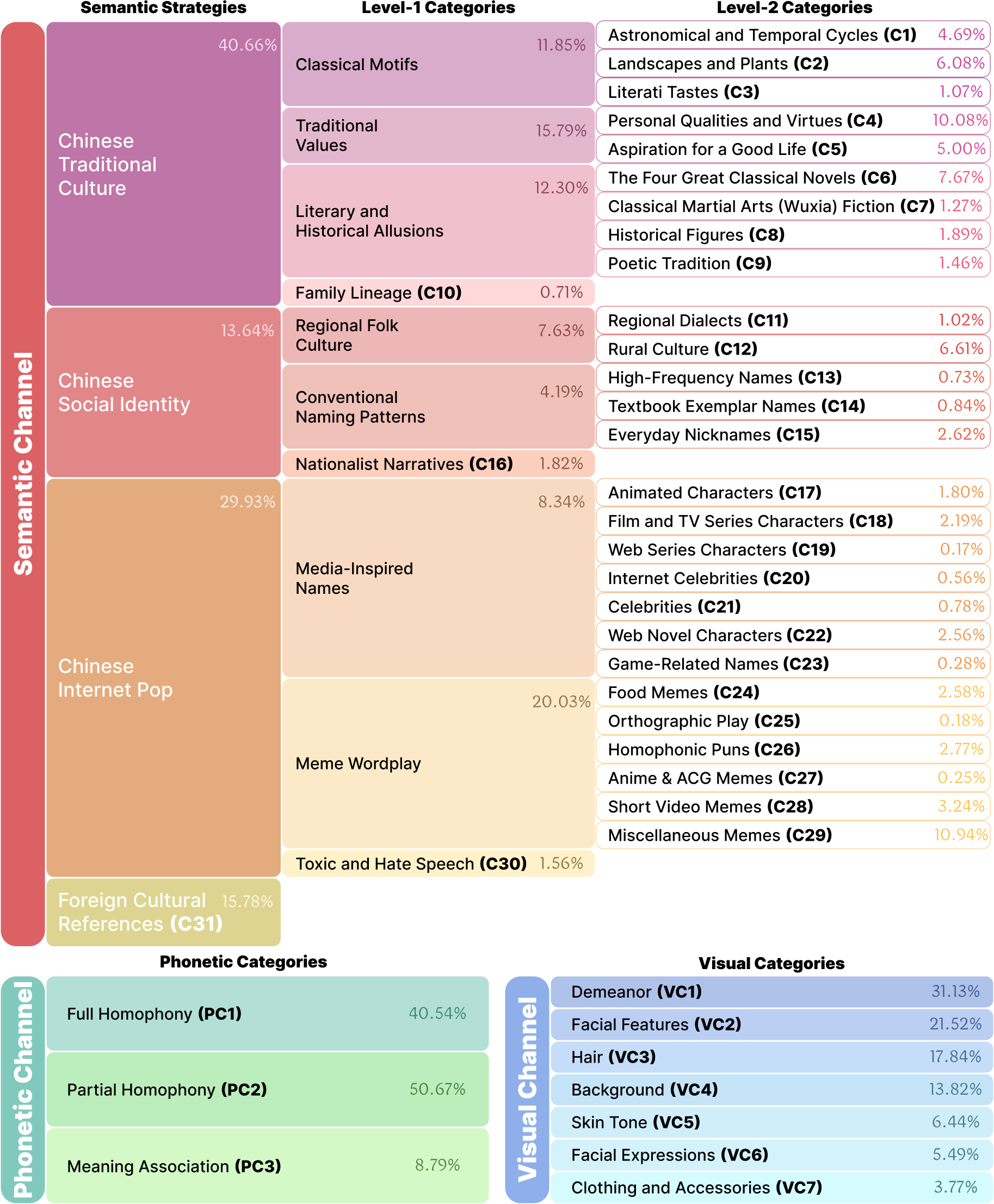}
    \caption{Framework of naming strategies across \textit{semantic}, \textit{phonetic}, and \textit{visual} channels. The semantic channel is organized hierarchically, including four overarching semantic strategies, ten Level-1 categories, twenty-seven Level-2 categories, and thirty-one subcategories at the most detailed level. The phonetic channel comprises three categories, while the visual channel includes seven categories.}
    \label{fig:Framework}
    \Description{Framework of naming strategies across semantic, phonetic, and visual channels. The semantic channel is organized hierarchically, including four overarching semantic strategies, ten Level-1 categories, twenty-seven Level-2 categories, and thirty-one subcategories at the most detailed level. The phonetic channel comprises three categories, while the visual channel includes seven categories.}
\end{figure*}

\subsection{A Framework of Cross-Cultural Naming Strategies}

\textcolor{black}{\autoref{fig:Framework}} shows our final framework of naming strategies consisting of three channels: semantic, phonetic, and visual.

\subsubsection{Semantic Channel}
Chinese encoders infused names with rich meanings with the following major strategies:

\textbf{Foreign Cultural References (15.78\%).} This strategy (C31) captures cases where Chinese users adapt their naming choices to align with the cultural background of foreign newcomers, thereby reducing social distance and converging toward the newcomers' culture \cite{gallois2005communication}. 
For example, names such as 乔·埃里克森 are recognized as ``foreign-style names'', with foreignness itself serving as the core meaning and signaling accommodation to newcomers' cultural identity. Similarly, names like 蒙娜丽莎 (Mona Lisa) function as cultural symbols, evoking associations with Western art and identity through well-known icons. However, the overall proportion of Foreign Cultural References remains relatively small (15.78\%) compared with Chinese-oriented strategies. This asymmetry suggests that although semantic convergence occurs, Chinese users are more inclined to impose their own cultural frameworks rather than fully accommodate foreign traditions.

\textbf{Chinese Traditional Culture (40.66\%).} 
This strategy mirrors common practices in domestic naming, giving the impression that foreigners are incorporated into Chinese cultural conventions. As the largest category, it reflects the dominance of Chinese cultural frames in cross-cultural naming practice. It represents a form of in-grouping on local cultural terms, where foreigners are assimilated into symbolic systems that define inclusion through Chinese traditions \cite{van1998ideology}.
This strategy can be divided into four first-level categories: \textbf{\textit{Classical Motifs}}, \textbf{\textit{Traditional Values}}, \textbf{\textit{Literary and Historical Allusions}}, and \textbf{\textit{Family Lineage}}. Names created through \textbf{\textit{Classical Motifs (11.85\%)}} project imagery from classical Chinese culture onto personal names. Such motifs are typically drawn from astronomical and temporal cycles (C1, e.g., sun, moon, stars, seasons), landscapes and plants (C2, e.g., pine, bamboo, jade, flowers), and literati tastes (C3, e.g., zizther, chess, and painting). For instance, the name 腊月 refers to the twelfth month of the traditional Chinese lunar calendar, symbolizing the end of the year and evoking seasonal and festive associations. \textbf{\textit{Traditional Values (15.79\%)}} represent naming strategies that embed Chinese notions of personal qualities and virtues (C4, e.g., integrity, loyalty, benevolence) or aspirations for a good life (C5, e.g., wealth, longevity, prosperity). For example, in the name 李富贵, 李 is the family surname, while 富贵 expresses the pursuit of wealth and social status. \textbf{\textit{Literary and Historical Allusions (12.30\%)}} are naming strategies that borrow from Chinese literary and historical traditions, including the four great classical novels (C6, e.g., Journey to the West, Dream of the Red Chamber), classical martial arts fiction (C7, e.g., The Heaven Sword and Dragon Saber), historical figures (C8, e.g., Qin Shi Huang), and poetic tradition (C9, e.g., The Book of Songs (Shijing)). For example, the name 黛玉 originates from \textit{Dream of the Red Chamber}, whose bearer is the novel's female protagonist.

\textbf{\textit{Family Lineage (C10, 0.71\%)}} is relatively rare and reflects Chinese traditions of ancestral heritage through compound surnames. For example, in the name 欧阳宁璇, 欧阳 is a traditional compound surname that traces back to aristocratic families of the Yue state in the Spring and Autumn period (770–476 BCE), carrying long-standing historical and cultural significance.

\textbf{Chinese Social Identity (13.64\%).} This strategy constitutes another form of divergence, but instead of drawing on elevated cultural conventions, it assigns newcomers to stereotyped social categories. This strategy resembles everyday labels that Chinese users associate with colloquial or archetypal identities. Such naming produces a form of as-if in-grouping, where foreigners are nominally included by being given Chinese-style names, yet the stereotypical labels simultaneously highlight their position as out-group members. This duality reflects the dynamics of social categorization \cite{tajfel2004social, gallois2005communication}. This strategy can be divided into three first-level categories: \textbf{\textit{Regional Folk Culture}}, \textbf{\textit{Conventional Naming Patterns}}, and \textbf{\textit{National Narrative}}. Among them, \textbf{\textit{Regional Folk Culture (7.63\%)}} is the most common form. It represents naming practices that draw on regional dialects (C11) or rural traditions (C12). For example, the name 狗蛋 (doggy) is a typical rustic name in traditional Chinese folk culture, originating from the custom of giving children ``lowly'' names under the belief that such names would help them survive and thrive. \textbf{\textit{Conventional Naming Patterns (4.19\%)}} represent naming strategies that convey stereotypical impressions by drawing on high-frequency names (C13), textbook exemplar names (C14), or everyday nicknames (C15). For example, 李华 is the most common fictional name in Chinese English-language exams.
\textbf{\textit{Nationalist Narrative (C16, 1.82\%)}} refers to naming strategies that embed patriotic narratives and collective ideals into personal names. For example, 建国 
was especially common among those born around the founding of the People's Republic in 1949 and literally means ``nation-building.''
When used for foreign newcomers, such names also convey patriotic sentiment, projecting national pride through the act of naming.

\textbf{Chinese Internet Pop (29.93\%).} This strategy is largely shaped by online trends and digital culture in contemporary China. Unlike strategies that create the impression of inclusion, Internet Pop often positions newcomers as outsiders through humor, parody, and playful exaggeration. Rather than integrating them into cultural or social conventions, these strategies turn foreigners into subjects of entertainment while also testing the boundaries of cross-cultural interaction \cite{gallois2005communication}. In this sense, Internet Pop operates as an out-grouping mechanism that reinforces difference through irony, teasing, and light-hearted play \cite{tajfel2004social,billig2005laughter,shifman2013memes}.
This strategy can be divided into three first-level categories: \textbf{\textit{Media-Inspired Names}}, \textbf{\textit{Meme Wordplay}}, and \textbf{\textit{Toxic and Hate Speech}}. \textbf{\textit{Media-Inspired Names (8.34\%)}} draw directly on popular media, including animated characters (C17), film and TV series characters (C18), web series characters (C19), internet celebrities (C20), celebrities (C21), web-novel characters (C22), and game-related names (C23). 
\textbf{\textit{Meme Wordplay (20.03\%)}} is the largest first-level category, encompassing names derived from internet memes and playful word transformations. Common categories include food-related memes (C24, e.g., names invoking pineapple pizza to parody Italians' aversion to it), homophonic puns (C25, e.g., ``蒂蒂薇'' (Dì Dì Wēi) created as a phonetic distortion of the pesticide brand ``敌敌畏'' (Dí Dí Wèi)), orthographic play (C26, e.g., rare or visually complex characters such as 龘龖龘 that exaggerate visual form rather than meaning), anime and ACG references (C27, e.g., combining Japanese anime clans with colloquial Chinese names for comic effect), and short-video tropes (C28, e.g., 小美, a formulaic character used in short-video narration). Because of the dynamic and participatory nature of Chinese internet culture, many memes circulate without clear origins, evolve rapidly, and resist precise categorization. We therefore introduced a miscellaneous meme category (C29) for name strategies that are widely recognized as playful memes but are too diffuse or hybrid to be assigned to a single subtype. For example, 死人但有点活 (``dead but somewhat alive'') is used as a name that exaggerates the state of exhaustion while still functioning, illustrating dark humor through ironic wordplay.
Finally,\textbf{\textit{Toxic and Hate Speech (C30, 1.63\%)}} refers to names that deliberately incorporate offensive or taboo expressions, including profanity, sexual references, or politically sensitive figures. 
Such strategies often function as out-grouping and reflect boundary-testing practices that reinforce stigma and social hierarchies \cite{tajfel2004social,gallois2005communication,van2017discourse,sidanius2001social}. For example, some names use crude references to body parts, such as 屌毛 (literally ``pubic hair''), while others revive historically racialized slurs such as 昆仑奴 (Kunlun slave), a derogatory term once used in China for dark-skinned servants from Southeast Asia or Africa.

\subsubsection{Phonetic Channel}
We also identified a set of naming strategies that link Chinese names to foreigners' original names through pronunciation. This includes full homophony, partial homophony, and meaning association (as classified in \autoref{sec-4.2.2}). 
Among the \textcolor{black}{62,461} data entries, \textcolor{black}{48,296} provided the original foreign names, accounting for 77.3\% of the dataset. Within these, 30.4\% employed naming strategies under phonetic channel. Given the availability of original names, Chinese users tended to reference them in the naming process, reflecting an orientation toward linguistic accommodation and cross-linguistic similarity \cite{giles1987ethnolinguistic} 

\textbf{Full Homophony (PC1, 40.54\%)} refers to naming strategies where the pronunciation of the Chinese name closely matches the foreign name, with only minor phonetic adjustments such as tone omission or consonant substitution. For example, 詹姆斯 (Zhān Mǔ Sī) is a fully phonetic transcription of the English name James. Each syllable in the Chinese rendering approximates a segment of the English pronunciation: ``詹'' (Zhān) corresponds to the initial /dʒeɪ/ sound, while ``姆斯'' (Mǔ Sī) approximates the coda /mz/.
\textbf{Partial Homophony (PC2, 50.67\%)} refers to naming strategies where a meaningful component of the Chinese name (e.g., a surname or a single character) phonetically matches part of the foreign name, while the remaining components are filled in through non-phonetic choices drawn from Chinese or Western repertoires. As the most frequent phonetic naming strategy, this pattern suggests that users often preferred to accommodate linguistically by preserving some phonetic similarity, while simultaneously embedding their own cultural or semantic preferences into the name. For example, 戴嘉诩 (Dài Jiā Xǔ) is a partial phonetic transcription of the English name Daniel. The surname ``戴'' (Dài) closely approximates the initial syllable /dæ/ in Daniel, while the remaining characters are not phonetically related to the original name.
\textbf{Meaning Association (PC3, 8.79\%)} does not rely on phonetic similarity but instead maps foreign names to Chinese names through direct semantic association. For example, 余烬 (Yú Jìn) is given for a foreign newcomer's name Ash. The Chinese word literally means ``embers'' or ``burning remains,'' which corresponds directly to the meaning of ash as the residue after fire.  

\begin{figure*}[h]
    \centering
    \includegraphics[width=\linewidth]{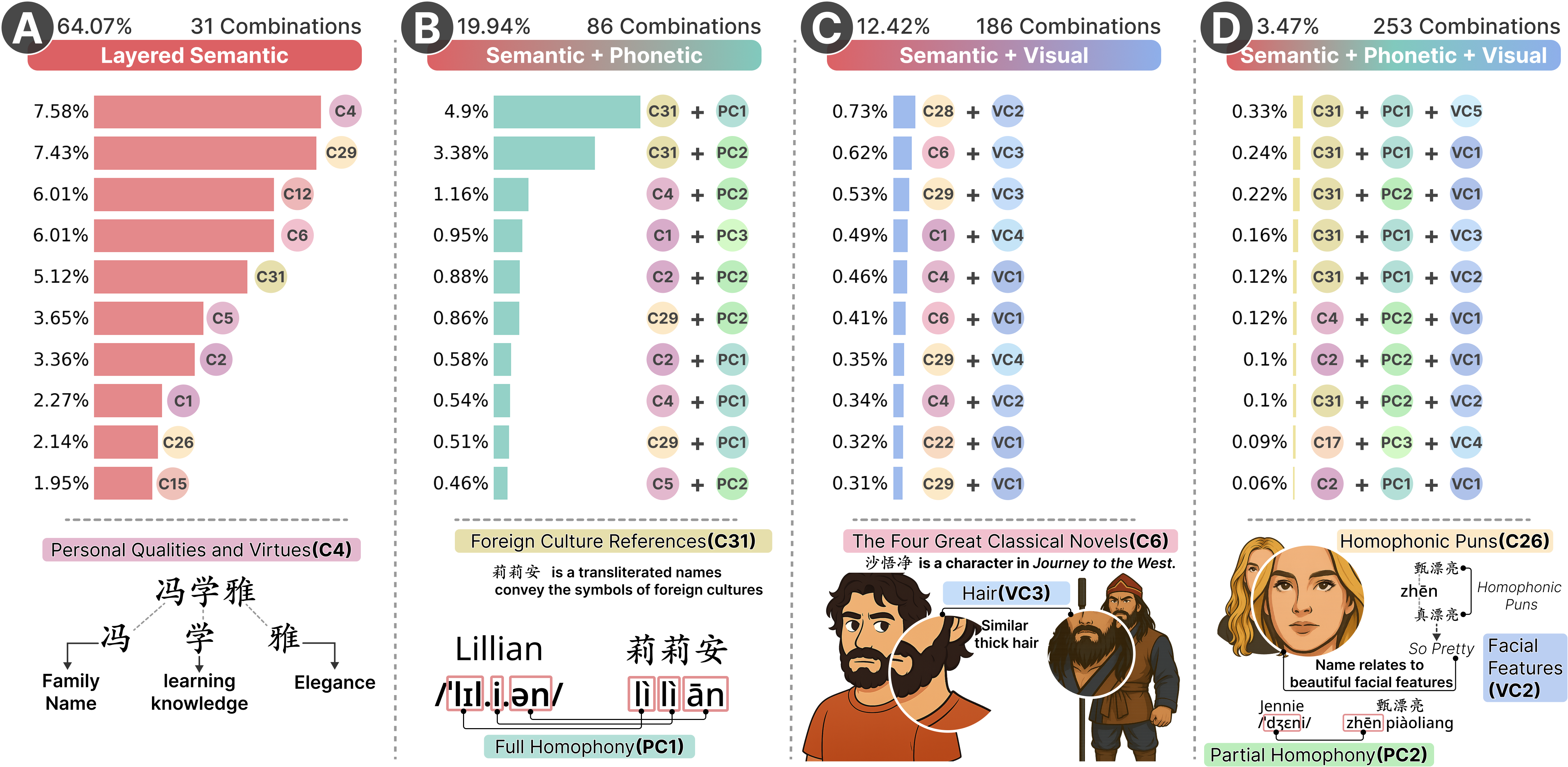}
    \caption{
    Top ten combinations within (A) Layered Semantic, (B) Semantic + Phonetic, (C) Semantic + Visual, and (D) Semantic + Phonetic + Visual, each with an illustrative example.
    }
    \label{fig:Layer}
    \Description{Top ten combinations within (A) Layered Semantic, (B) Semantic + Phonetic, (C) Semantic + Visual, and (D) Semantic + Phonetic + Visual, each with an illustrative example.}
\end{figure*}

\subsubsection{Visual Channel}
The visual channel captures how Chinese names are associated with foreigners' photos, including demeanor, facial features, hair, background, skin tone, facial expressions, and clothing and accessories (as classified in \autoref{sec-4.2.2}). Among the \textcolor{black}{62,461} data entries, \textcolor{black}{48,310} (77.3\%) included profile photos, and 20.67\% of these showed visual links between the assigned name and the poster's photo. This suggests that visual cues played a notable role in the naming process, as Chinese users drew on observable physical attributes when generating names for foreign newcomers.

Specifically, \textbf{Demeanor (VC1, 31.13\%)} refers to the overall impression of attitude or aura conveyed in the photo (e.g., relaxed, stern). \textbf{Facial Features (VC2, 21.52\%)} indicate specific physical characteristics of the face (e.g., eyes, nose, mouth). \textbf{Hair (VC3, 17.84\%)} covers head or facial hair, including its color and style (e.g., hair, beard). \textbf{Background (VC4, 13.82\%)} points to the visible setting behind the person (e.g., wild, kitchen). \textbf{Skin Tone (VC5, 6.44\%)} denotes the visible shade of the skin (e.g., light , dark). \textbf{Facial Expressions (VC6, 5.49\%)} describe the expressions shown on the face (e.g., smiling, frowning). \textbf{Clothing and Accessories (VC7, 3.77\%)} highlight visible garments or adornments (e.g., glasses, hats, earrings). For example, in one foreign newcomer's photo, the person had very short, light blond hair. Chinese users named him ``黄毛'' (literally ``yellow hair''), creating a direct visual correspondence between the blond hair and the literal meaning of the name.

\subsection{How the Strategies are Combined}
\label{sec-5.3}

\textcolor{black}{
Drawing on prior literature, we propose three hypotheses to guide our analysis of how these channels are combined:}

\textcolor{black}{First, research on the ideographic nature of Chinese characters has argued that written forms operate primarily as meaning-bearing units \cite{zhang2024ideograph}, and studies on Chinese naming have shown that personal names are expected to convey layered intentions such as virtues, aspirations, or symbolic associations \cite{gao2011shall}. Given this tradition and the dominance of semantic strategies in our dataset, we posed \textbf{H1: In channel combinations, multi-layered semantic channels will be prioritized}.
Second, work on name translation has indicated that cross-language adaptation often relies on phonetic approximation, while semantic refinement is favored when culturally acceptable \cite{kim2020linguistics}. Thus, we posed \textbf{H2: In dual-modal encoding, phonetic channels tend to be favored.}
Third, research on proper names in multimodal discourse has shown that names in the linguistic landscape frequently draw on several semiotic resources simultaneously, requiring coordinated interpretation across written form, sound, co-occurring visual elements and generating structural complexity \cite{sandst2020proper}. Based on this, we posed \textbf{H3: The frequency of names will be highest when encoded through a single channel and will decrease as more channels are added.}
}

\textcolor{black}{With these questions in mind, we examined how channels are used in combination in our dataset (the results are summarized in \autoref{fig:Layer}).}
\textcolor{black}{To test H1, we conducted a chi-square test comparing the distribution of instances across channel combinations. The test revealed a significant deviation from a uniform distribution (χ²(3) = 54,208.67, p < .001), with a large effect size (Cramér's V = 0.54), indicating a strong dominance of semantic-only channels.}
\textcolor{black}{Consistent with this result, we found that 64.07\% of names used multi-layered semantic strategies, far exceeding the frequency of other combinations and making it the dominant pattern. This dominance confirms that meaning-oriented strategies play the primary role in the Chinese naming process. Thus, \textbf{H1 is supported}.} As shown in \autoref{fig:Layer} A, the most used semantic patterns are Personal Qualities and Virtues (C4), Miscellaneous Memes (C29), and Rural Culture (C12).
For example, the name 冯学雅 combines two distinct semantic elements: 学, meaning study, and 雅, meaning elegance or being well-educated. \textcolor{black}{Both elements speak to the traditional Chinese appreciation for scholarly pursuit and moral character.}

Dual-modal encodings account for 32.36\% of all cases.
\textcolor{black}{To test H2, we compared dual-modal encodings involving semantic–phonetic and semantic–visual combinations. A chi-square test revealed a significant imbalance between the two conditions (χ²(1) = 1091.16, p < .001), with a moderate effect size (Cramér's V = 0.23), indicating a preference for semantic–phonetic combinations over semantic–visual ones.
}
\textcolor{black}{Within these cases, semantic–phonetic combinations represent the largest share at 19.94\%, exceeding the proportion of semantic–visual combinations (12.42\%) thus \textbf{supporting H2}.}
\textcolor{black}{\autoref{fig:Layer} B shows frequent pairs of semantic–phonetic combinations. For example, some names reproduce the foreign pronunciation through commonly used transliteration characters,} as in 丽莲 (Lì lián), where sound corresponds to ``Lillian'': ``丽'' (lì) approximates ``Li'' and means beauty, while ``莲'' (lián) imitates ``lian'' and refers to the lotus, a cultural symbol of purity.
戴嘉诩 derives from ``Daniel'': ``戴'' (dài) approximates ``Dan,'' while ``嘉'' (jiā) means excellence and ``诩'' (xǔ) means praise, adding laudatory meanings that go beyond sound imitation. 
\textcolor{black}{Compared with these high-frequency patterns, semantic–visual combinations (\autoref{fig:Layer} C) generally show low frequency, emerging primarily when visual traits reinforce stable associations with specific characters. For example, one user named a bearded foreign man 沙悟净. This name originates from the classic Chinese novel Journey to the West and immediately resonates with the image of a man with thick beard.}

\textcolor{black}{To assess H3, we examined whether name frequency decreases as more encoding channels are combined, using a Negative Binomial regression. The number of channels showed a significant negative effect (β = −1.44, p = .041), indicating that frequency decreases as more channels are combined; each additional channel reduced the expected frequency to about 24\% of its previous level (exp(β) = 0.24).}
\textcolor{black}{As the number of multimodal channels increases, the frequency of names correspondingly decreases. Names utilizing all three channels—semantic, visual, and phonetic—constitute only 3.57\% of all cases. This contrasts sharply with single-channel strategies (64.07\%) and dual-channel ones (32.36\%), making it the rarest pattern in our data (\autoref{fig:Layer} D), a finding that \textbf{supports H3}.
These cases require the simultaneous satisfaction of multiple conditions: phonetic resemblance, cultural meaningfulness, and alignment with the visual features or impressions of the foreigners being named.} For example, the name 甄漂亮 (Zhēn Piàoliang) is derived from the English name ``Jenny,'' using ``甄'' (zhēn) as a partial phonetic echo and combining it with the evaluative term ``漂亮'' (beautiful), a reading reinforced by the commenter's interpretation of the user's appearance. Furthermore, 甄漂亮 is itself a popular internet meme among young Chinese, often used to refer to characters in detective shows, adding another layer of contemporary cultural reference.
\textcolor{black}{To examine potential interdependencies among channels, we assessed multicollinearity using variance inflation factors (VIFs) computed on an instance-level channel presence matrix. All VIF values were low (semantic: 1.50; phonetic: 1.00; visual: 1.00), indicating no severe multicollinearity.}

\textcolor{black}{Together, these combinations assemble a diverse, multimodal array of cultural resources into a complex digital ``Babel tower'' that challenges cross-cultural communication.}

\subsection{Engagement Analysis}
\label{sec-5.4}
\textcolor{black}{Lastly, we analyzed engagement patterns in these cross-cultural naming practices using the number of likes as a proxy for virality, thereby identifying which naming strategies performed better on RedNote. Besides, like most social media, comments on RedNote are sorted in reverse order of likes, meaning that comments with more likes are more visible to foreign newcomers. This sorting logic, therefore, shapes both the visibility structure of, and the priority within, the ``Babel tower.''
}

\begin{figure*}[h]
    \centering
    \includegraphics[width=\linewidth]{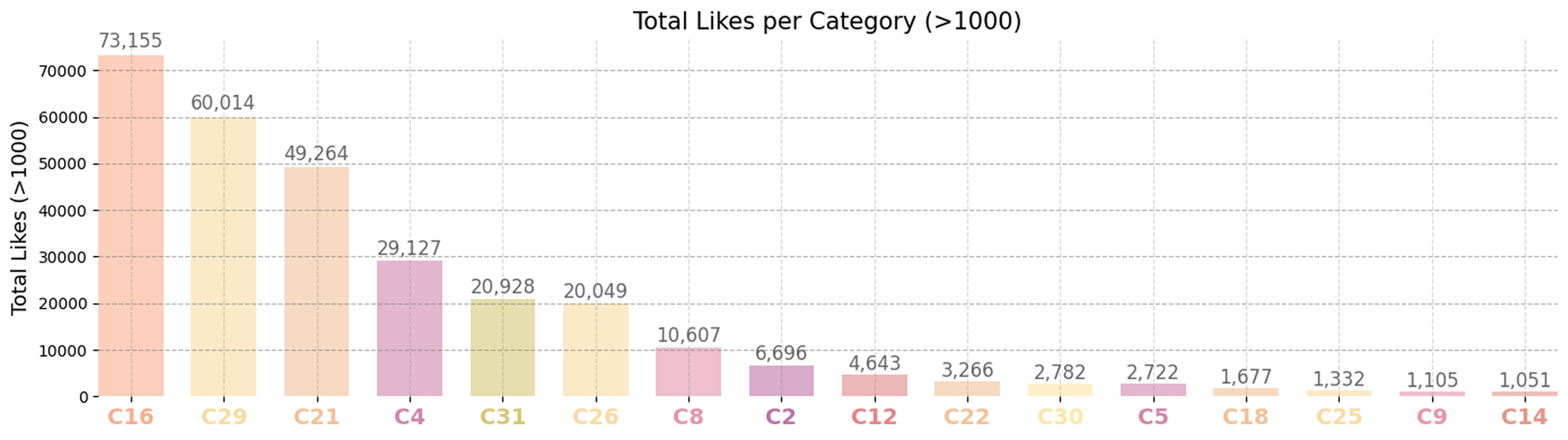}
    \caption{%
    Distribution of viral likes (≥1000) across categories.
    }
    \label{fig:Analysis}
    \Description{Distribution of viral likes (≥1000) across categories.}
\end{figure*}

Overall, the distribution of likes shows a strong zero-inflation pattern: the vast majority of comments receive no likes, consistent with the long-tail dynamics of social media. To better capture influence, we focused on comments with viral likes (\textcolor{black}{\autoref{fig:Analysis}}). In this ranking, National Narratives (\textbf{\textit{C16}}, 73,155 likes), Miscellaneous Memes (\textbf{\textit{C29}}, 60,014 likes), and Celebrities (\textbf{\textit{C21}}, 49,264 likes) stand out above other categories.
\textcolor{black}{Compared with \autoref{fig:Framework}, the most significant discrepancy lies in the fact that the category with the largest actual proportion in the dataset (e.g., Chinese Traditional Culture) was not the most viral. Conversely, categories that occupied only minor shares in the framework—such as National Narratives (e.g., assigning foreigners names with socialist or nationalist connotations) and various internet memes and celebrity references—emerged as the most-liked categories.}

\textcolor{black}{This intriguing discrepancy, particularly the virality of parody and degradation, resonates with Bakhtin's \textbf{theory of carnival}. The theory posits that carnival represents a temporary interactional order in which ordinary hierarchies loosen and participants engage with one another through \textbf{parody, humbling gestures, and coarse expression} \cite{bakhtin2013problems}. 
In the context of our study, parody operates through exaggeration, appropriation, and imitation, inserting the foreign newcomers into narratives drawn from Chinese popular culture and internet memes. For instance, the highly upvoted names ``张艺兴'' (C21, a well-known Chinese pop singer, 34{,}237 likes) and ``张大大'' (C21, a Chinese well-known celebrity, 15{,}027 likes) do not imply that the newcomer is literally a star. Rather, they borrow celebrity templates to produce an entertainment oriented parody, playfully turning an unfamiliar foreign individual into a recognizable character within Chinese online culture. Degradation, on the other hands, pulls what is elevated down into the everyday or even the ridiculous. A representative case is ``A大英国艺术品零售批发老张'' (C29, 22{,}845 likes), a meme-style name modeled after labels used by online resellers. The original English name \textit{Art}, which carries connotations of artistic refinement, is placed in a mundane commercial setting and combined with the very local label ``老张.'' This downward movement rewrites the foreign newcomer's identity from refined to ordinary and from distant to close, offering a classic example of carnival-style degradation.} 
\textcolor{black}{These names adopt parody and degradation that largely rely on meme-based humor, playful riffs, and light teasing. Such humor reflects a familiar mode of engagement \cite{bakhtin2013problems}, reducing social distance rather than signaling a formal cross-cultural exchange. 
}

\textcolor{black}{Another core feature of carnival, coarse expression \cite{bakhtin2013problems}, is also present in our corpus but with few likes. Toxic and Hate Speech (\textbf{\textit{C30}}, 1.56\% of comments, 2,782 in total), including clearly insulting items such as ``昆仑奴'' (historically used to describe dark-skinned servants from Southeast Asia or Africa, 2 likes), attracts very few likes, suggesting that overtly hostile language remains marginal, either because users are reluctant to endorse it or because platform moderation limits its visibility. By contrast, seemingly coarse but locally affectionate nicknames such as ``狗蛋'' (C12, a ``lowly'' childhood nickname sometimes used in rural families to symbolically protect a child, 527 likes) circulate much more widely: while they may appear demeaning from a cross-cultural perspective, many Chinese users perceive them as playful or endearing. As a result, the content most likely to be misread across cultures is not the most explicitly toxic material, but these ambiguous, carnivalesque terms; we return to this issue in the Discussion section.}

Taken together, these engagement patterns highlight the carnivalesque ways Chinese users engage with foreign newcomers. 
\textcolor{black}{Comments with viral numbers of likes suggest that foreign newcomers are no longer treated as solemn cultural others but as textual resources that can be rewritten, appropriated, and made playful, while still conveying a warm and familiar form of engagement. These naming forms spread widely because they align with user practices focused on entertainment, are amplified by platform algorithms, and benefit from cumulative like dynamics that push them to the top of comment threads, making them especially visible to foreign newcomers. Within this dynamic, some forms, such as Traditional Culture, circulate broadly across the corpus, whereas others, such as Memes and Nationalist Narratives, attract disproportionate numbers of likes and are more likely to be surfaced. Together, these naming strategies, their distributions, and their differential visibility to foreign newcomers constitute the final presence of this ``Babel Tower'' of cross-cultural communication.}

\section{Discussion}
\label{sec-6}
This section outlines the study's implications in terms of methodology, cross-cultural communication, and platform design, \textcolor{black}{as well as its limitations and future directions.}

\subsection{Methodological Implications for Cross-Cultural Data Analysis}
This study highlights methodological challenges in cross-cultural social media analysis and offers lessons for future work.

\textbf{Human-in-the-Loop Correction for Scalable Extraction via LLM Crowdsourcing.} Cross-cultural social media data are typically massive, noisy, and semantically diverse. To handle this, we operationalized extraction by treating multiple LLMs as a synthetic crowd and applying consistency-guided human correction. This human-in-the-loop design reduced annotation cost, required minimal additional time, and improved accuracy.
In sum, LLM-as-crowd with consistency-based arbitration offers a portable template for extracting structured information from messy cross-cultural corpora and can be adapted to other domains facing similar challenges.

\textbf{Leveraging LLMs for Cross-Cultural Interpretation and Generation} A further challenge was that many records contained missing or partial explanations. In cross-cultural communication data, such incomplete cases are often highly complex and unusual, involving not only cultural and linguistic nuances but also internet memes, homophonic puns, and even visual reasoning. Even setting aside the immense time investment, these dimensions are extremely difficult for human coders to interpret, as they require a vast reservoir of knowledge spanning multiple domains. LLMs enhanced efficiency in handling such cases and provided cultural and symbolic knowledge that extended beyond human expertise. Human coders then played a critical role in iterative validation, identifying common error patterns and codifying them into correction heuristics. This paradigm shift enabled the processing of cross-cultural data at a scale and level of reliability unattainable in traditional studies.

\textbf{Future Agenda.} Despite these advances, current AI models remain insufficient for fully capturing the interplay of cultural semantics, phonetics, and visuals. In practice, they sometimes fail to extract content that is obvious to humans. \textcolor{black}{For example, in a comment where users separately explained that ``安'' is a Chinese family name and ``娜'' describes a pretty girl, human coders can easily infer that the intended name is ``安娜''. LLMs, however, often misidentify such cases when semantic clues are distributed across multiple short responses rather than presented in a single coherent statement.} Models also struggle to judge meanings of some internet memes and humor within the unique contexts of social media. \textcolor{black}{For instance, ``齐德龙'' (C29) comes from the comedic phrase ``齐德隆咚锵,'' an internet meme based on a well-known sketch in China. Human coders can easily recognize this playful reference, but the LLM treated ``齐'' as a common surname and interpreted ``德龙'' as virtue and auspiciousness (C4), resulting in a conventional, literal interpretation. In such cases, the explanation is not strictly wrong, but it misses the intended humorous or social function.} 
\textcolor{black}{
Thus, during the human review process used to verify and correct LLM-generated explanations for a subset of data in \autoref{Sec-4.2.3}, we observed that while all performance metrics were generally strong, the LLM's performance on contextual appropriateness was slightly lower than others. This suggests that capturing contextual and connotative meaning at scale remains challenging, particularly in high-context and rapidly evolving social media environments.}
Finally, models occasionally generate hallucinations or divergent results even under clear rules. \textcolor{black}{For example, when generating phonetic explanations, the model judged ``张天合 (Zhāng Tiānhé)'' to involve partial homophony (PC2) by claiming that ``合 (hé)'' corresponds to the English name ``Art'' (/ɑːrt/), even though the two share no phonetic correspondence.} Limited human adjudication cannot compensate for these gaps at scale. 
Together, these challenges underscore the need for culturally enriched model training that can better support transformation, association, and inference in diverse contexts. Our approach demonstrates what is achievable with current naming practice data while also pointing to open opportunities for further innovation. As cross-cultural platforms proliferate, such human-in-the-loop strategies will become increasingly important.

\subsection{Implications for Cross-Cultural Communication and Platform Design }
This study shows how multichannel naming practices complicate cross-cultural dialogue and points to implications for communication theory, platform governance, and technology design. 
 
\textbf{Extending High-Context Communication to Digital Environments.} Our findings show that multichannel naming practices function as a cultural encoding strategy that intensifies the dynamics of high-context communication in online spaces. Following Hall's definition~\cite{hall1976beyond}, in high-context cultures much of the information is conveyed through shared knowledge and situational cues rather than explicit linguistic content. On semi-anonymous platforms, users can deliberately construct additional contextual layers that not only transmit meaning but also reinforce group identity and delineate symbolic boundaries. 
Playful names \textcolor{black}{such as ``张艺兴'' (C21) or ``A大英国艺术品零售批发老张'' (C29)} create a carnivalesque arena \textcolor{black}{through parody and degradation where insider humor becomes highly visible. \textcolor{black}{Within this arena, interaction is experienced less as ordinary communication than as play: participants temporarily set aside the norms and expectations of their everyday lives, engaging with others from otherwise distant cultural backgrounds in a shared moment of improvisation and enjoyment.} These names receive strong support and are voted to the top, making them more prominent for newcomers to encounter (as discussed in \autoref{sec-5.4}).} Participation in this space requires sensitivity to fast-changing online humor. More formal names draw on historical or moral symbolism to assert cultural legitimacy. \textcolor{black}{For example, ``腊月'' (C1), which refers to the twelfth month of the traditional Chinese lunar calendar, also require specific cultural knowledge to decode.}
\textcolor{black}{When such cultural cues are unfamiliar, the underlying codes behind names may remain opaque to foreign newcomers, creating confusion or mismatched expectations.} Digital high-context communication therefore not only preserves implicitness but also creates thresholds of inclusion and exclusion in cross-cultural interaction.

\textbf{Cross-Cultural Carnival and the Governance of Playful Encounters.} 
The internet carnival observed in our case serves as a unique and potent mode of serendipitous cross-cultural communication. This phenomenon is collective, meme-driven, and fundamentally playful. Unlike structured cultural exchange, this low-commitment, game-like interaction allows users to temporarily set aside formal norms and engage across linguistic and cultural boundaries through humor, parody, and shared cultural references. However, this very spontaneity presents a dual governance challenge for platforms. 
First, it triggers a classic community growth dilemma. As suggested by prior literature \cite{kiene2016surviving}, sudden influxes of newcomers can disrupt a platform’s existing social ecosystem, forcing it to rebuild norm stabilization and support long-term assimilation. Second, and more specific to this cross-cultural context, the carnival introduces an acute communicative risk. The rapid, humor-laden exchange greatly amplifies the coding-decoding difficulty, where playful ambiguity can inadvertently lead to cultural exclusion or even toxic speech when jokes misfire or meanings are weaponized.
Addressing these intertwined challenges requires platform governance to prioritize interpretive flexibility over binary rulings. For instance, culturally layered content could be flagged as requiring context, enabling communities to crowdsource clarification \cite{gaver2003ambiguity}. When intervention against genuine harm is necessary, low-penalty actions paired with public explanations can educate users and reduce secondary conflict \cite{jhaver2024bystanders}. Concurrently, to support long-term integration, platforms should invest in socio-technical infrastructure, such as context-aware translation, that helps convert episodic bursts into sustained engagement (see the next point).
Ultimately, the goal is not to suppress the carnival but to architect participatory systems that nurture its connective potential while thoughtfully navigating the disruption and misunderstanding it may bring.

\textbf{From Semantic to Cultural Translation.} Mainstream machine translation systems and LLMs are still limited because they prioritize mapping words rather than interpreting intent. As a result, they often fail to capture the layered play of memes, puns, and symbolic associations that circulate in cross-cultural settings. 
\textcolor{black}{For example, the machine translation of the name ``翠花'' as ``Jade Flower'' captures only its literal meaning and misses the cultural associations in both rural naming traditions and contemporary humorous usage (\autoref{sec-5.3}).} Our findings point to the need for translation to move beyond semantic equivalence toward cultural translation. This requires models trained not only on standardized corpora but also on community-specific and multimodal data, capable of integrating phonetic, semantic, and visual cues. More importantly, systems must be designed to recognize communicative purpose: when humor, parody, or symbolism is intended, a literal rendering is insufficient. Translation that accounts for cultural intent as well as linguistic form would produce outputs that are not only intelligible but also resonate across cultural boundaries. 
In this way, AI systems can evolve from being simple language converters to becoming interpretive bridges in cross-cultural communication. 

\textbf{Future Agenda.} Despite these insights, important questions remain. Our analysis focused on a single platform, RedNote, \textcolor{black}{whose platform affordances and the unusually large influx of Western newcomers during the ``TikTok refugee'' event co-shaped how these naming practices unfold.}
\textcolor{black}{Future work should examine how similar or contrasting platform affordances, for example the video-centred formats of YouTube and TikTok or the interaction patterns found on X, influence the emergence and circulation of such naming strategies across cross-cultural digital settings.}
\textcolor{black}{Multichannel cultural encoding involving phonetic, semantic and visual cues frequently appears in digital interaction across platforms, and insider and outsider meaning dynamics often shape cross-cultural encounters. Misaligned offence and cross-cultural confusion are also common features of multilingual and multicultural online communication.} Future research should examine multiple platforms and cultural contexts to understand how these strategies emerge and circulate under different media affordances and community compositions. As AI-based moderation and translation tools become more widely adopted, the tension between cultural authenticity and cross-cultural accessibility will become even more pronounced. Future work should investigate not only how insiders construct cultural boundaries but also how outsiders adapt, negotiate, or resist them. Expanding the scope from naming to other symbolic practices such as memes, images, and multimodal discourse would deepen our understanding of digital cross-cultural interaction.

\subsection{Limitations}
First, our tag-based and engagement-filtered sampling captures active naming discussions but may bias the dataset toward more visible posts. Future work could explore complementary sampling strategies, such as semantic retrieval or stratified sampling across engagement tiers, to obtain a more comprehensive view of cross-cultural naming practices. In addition, RedNote's platform restrictions limit our dataset to first-level comments and like counts, preventing reconstruction of deeper reply chains or identification of commenter-level interaction patterns. As a result, our analysis centers on direct naming responses and observable signals of visibility and endorsement rather than full conversational trajectories. Future research could incorporate richer conversational logs, multi-level reply structures, or audience-level metadata to more fully trace cross-cultural interactions and multi-level decoding processes.


\vspace{1em}
\section{Conclusion}
\label{sec-7}

This paper examined cross-cultural naming practices on RedNote, where Chinese users assigned Chinese names to foreign newcomers during the TikTok refugee event. To analyze this phenomenon, we developed a creative human–AI collaborative approach and constructed a systematic three-channel framework of naming strategies. Our analysis demonstrates how semantic, phonetic, and visual channels combine to create layered forms of meaning that complicate interpretation, and how naming strategies are enacted in practice. Taken together, these findings show how the interplay of naming strategies construct a  mordern ``Babel Tower'' in cross-cultural communication and provide implications for platform design aimed at supporting effective cultural exchange in global social-media environments.

\begin{acks}
This work was supported by National Natural Science Foundation of China (No.62402121, No.62472099), Shanghai Chenguang Program, Research and Innovation Projects from the School of Journalism at Fudan University, and Cyrus Tang Foundation.
\end{acks}

\bibliographystyle{ACM-Reference-Format}
\bibliography{main}

\end{CJK*}
\end{document}